%% file: battery_asymptotics_COMBINED.tex
\documentclass[final,onefignum,onetabnum]{siamart171218}
\input{battery_asymptotics_shared}

\ifpdf
\hypersetup{
	pdftitle={Asymptotic reduction of a porous electrode model for lithium-ion batteries},
	pdfauthor={I. R. Moyles, M. G. Hennessy, T. G. Myers, and B. R. Wetton}
}
\fi

\begin{document}
	
	\maketitle
	
	% REQUIRED
	\begin{abstract}
		We present a porous electrode model for lithium-ion batteries using Butler--Volmer reaction kinetics.  We model lithium concentration in both the solid and fluid phase along with solid and liquid electric potential.  Through asymptotic reduction, we show that the electric potentials are spatially homogeneous which decouples the problem into a series of time-dependent problems.  These problems can be solved on three distinguished time scales, an early time scale where capacitance effects in the electrode dominate, a mid-range time scale where a spatial concentration gradient forms in the electrolyte, and a long-time scale where each of the electrodes saturate and deplete with lithium respectively.  The solid-phase concentration profiles are linear functions of time and the electrolyte potential is everywhere zero, which allows the model to be reduced to a system of two uncoupled ordinary differential equations. Analytic and numerical results are compared with full numerical simulations and experimental discharge curves demonstrating excellent agreement. 
	\end{abstract}
	
	% REQUIRED
	
	% REQUIRED
	
	\section{Introduction}
	
	Rechargeable lithium-ion batteries (LIBs) are ubiquitous in society, being utilised in medical devices, mobile phones, and transportation vehicles such as cars and airplanes.  LIBs currently dominate the energy storage market compared to other batteries mostly due to a long lifetime, high energy densities, and low self-discharge rates \cite{Tarascon2001}.  As society moves to lessen the demands on traditional energy sources and increase the demands of portable electronics, higher capacity and safer LIBs are required.
	
	Experimental studies are crucial in improving battery performance and lifetime \cite{Kennedy2017, Kennedy2014, Stokes2017}.  However, battery prototypes are expensive to produce since a large number of experiments are required to assess the impact of new designs.  Mathematical modelling can alleviate this pressure by providing a means to identify, simulate, and simplify dominant physics in battery operation at a fraction of the cost.
	
	Since the seminal work of Newman \cite{Newman1962}, who pioneered continuum modelling of porous electrochemical batteries, a plethora of works have appeared that address mathematical models and their simulation to a varying degree of complexity. A full review of these results is outside the scope of this manuscript; however, recent overviews can be found in Refs.~\cite{Gomadam2002, Ramadesigan2012}. Generally, theoretical developments follow three categories: (i) improved physical and electrochemical modelling \cite{Dargaville2010, Doyle1993, Farrell2000, Fuller1994, Johansen2006, Johnson1971, Newman1975, Ong1999, Pinson2013, Smith2017,Smith2017b,Biesheuvel2010,Biesheuvel2011,He2018,Mirzadeh2014,Singh2018}, (ii) analysis of mathematical models \cite{Richardson2012, Richardson2007} and (iii) large-scale model simulation \cite{Amiribavandpour2015, An2018,Li2014, Safari2011}.

	Articles in (i) focus on modelling new electrochemical and physical processes or improving current models.  This involves modelling capacitance processes \cite{Ong1999,Biesheuvel2010,Biesheuvel2011,He2018,Mirzadeh2014}, intercalation kinetics \cite{Smith2017b,Singh2018}, active-material utilisation \cite{Dargaville2010}, mechanics \cite{Chakraborty2015, Foster2017}, phase separation \cite{Smith2017, Ferguson2012, Ferguson2014, Orvananos2014}, and applying modelling results to commercial batteries.  While these models often advance the understanding of battery physics, they can be cumbersome to solve and may not elucidate dominant processes during battery operation. Articles in (ii) which address model analysis have considered the asymptotic reduction of homogenised battery models in the limit of small lithium concentration in the open-circuit potential \cite{Richardson2012} and also derived appropriate Butler-Volmer boundary conditions using matched asymptotic expansions \cite{Richardson2007}.  This approach attempts to identify the equations in a model which are most responsible for an observed behaviour, but sometimes requires unrealistic parameter values or leads to conclusions which cannot be related to practical batteries.  Large-scale simulations in (iii) tend to focus on adding complexities to simple models and studying the results.  These include using concentrated solution theory for the electrolyte, including temperature and compositional dependence in model parameters, and introducing different modelling domains for the solid and liquid phases.  This approach tends to better address battery practicality since realistic battery parameters and geometries can be utilised. Large-scale simulations tend to be computationally expensive and implemented in commercial software; however, optimised algorithms built on state-of-the-art routines can reduce some of the computational challenges.
	
	The aim of this paper is to bridge the areas of modelling, analysis, and simulation by performing a systematic asymptotic reduction of a practical model of LIBs.  The model is similar to that derived by Newman \etal\cite{Newman2004, Newman1975} using porous electrode theory and utilised by An \etal\cite{An2018}, Li \etal\cite{Li2014}, and Amiribavandpour \etal\cite{Amiribavandpour2015} to study the behaviour of commercial LIBs.  The simulation results of the latter two papers indicate that concentration profiles quickly settle into a steady state or evolve linearly with time and we will systematically show how this occurs. We compare to experimental results of Li \etal and show excellent agreement.  
	
	The paper is organized as follows. We summarise the non-dimensional volume-averaged porous electrode model in \cref{sec:model}.  We state an asymptotically reduced LIB model in \cref{sec:asymptotics} and derive it by exploiting the smallness of dimensionless parameters. We show how the asymptotic analysis admits analytical solutions valid in a series of time regimes which describe the entire battery discharge process.  The asymptotic solutions are compared against numerical simulations in \cref{sec:comparison} and battery discharge data in \cref{sec:data}. A discussion of the results follows in \cref{sec:results} and the paper concludes in \cref{sec:conclusions}.

	\section{Model overview}\label{sec:model}
\begin{figure}
		\centering
		\includegraphics[scale=0.7]{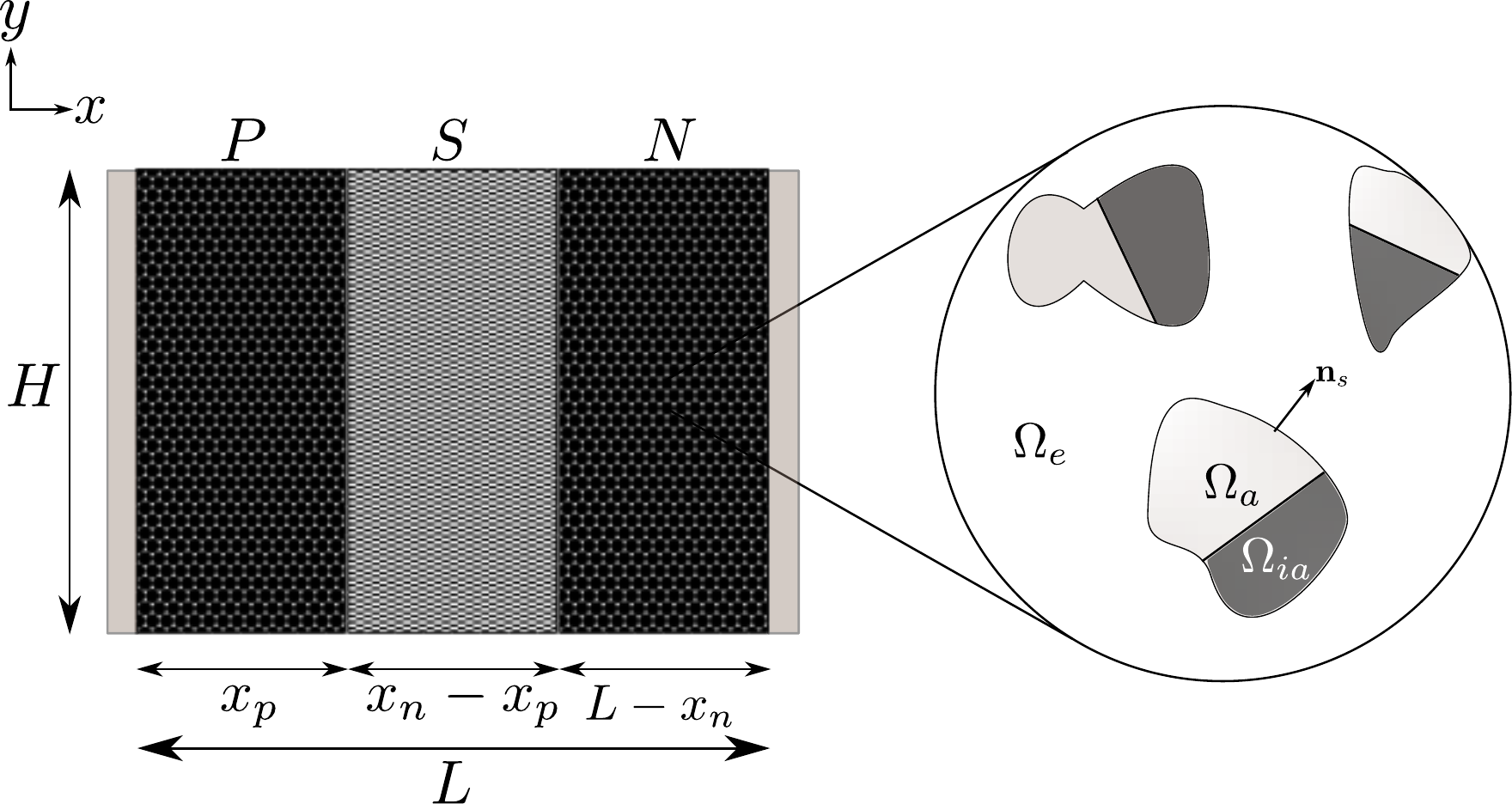}
		\caption{Setup of a battery cell including the porous structure of the electrode layers.}
		\label{fig:setup}
	\end{figure}

	We consider the electrochemical processes that occur in a single cell of an LIB, as shown in \cref{fig:setup}. The cell is composed of a positive ($P$) electrode, a separator ($S$), and a negative ($N$) electrode. The cell is assumed to be two dimensional with length $L$ and height $H$. The horizontal and vertical coordinates $x$ and $y$ are used to describe material points within the cell. The positive electrode exists on $0\leq x\leq x_p$, the separator on $x_p\leq x\leq x_n$, and the negative electrode on $x_n\leq x\leq L$. 
	
	The electrodes are porous and filled with an electrolyte that is able to carry ionic charge but not electrons. The solid material of each electrode contains active and inactive components. The active material carries electrons and hosts intercalated lithium which release as ions into the electrolytic phase. A typical electrode volume element can be decomposed into three subdomains corresponding to the active ($\Omega_\text{a}$) and inactive ($\Omega_\text{ia}$) materials and the void space occupied by the electrolyte ($\Omega_\text{e}$); see \cref{fig:setup}. The positive electrode lithiates on discharge and for this reason we assume it undergoes a chemical reaction of the form
	\begin{align}
	\ce{LiX
		<=>[charge][discharge]
		\ce{X} + \ce{Li^+} + \ce{e^-}
	},
	\end{align}
	where $X$ is a binding agent such as  \ce{CoO2}, \ce{Mn2O4}, \ce{FePO4}, and \ce{NiO2} \cite{Wang2012}. 
	Similarly, for the negative electrode, which delithiates on discharge, we assume a reaction of the form
	\begin{align}
	\ce{Y + \ce{Li+} + \ce{e-}
		<=>[charge][discharge]
		\ce{LiY}
	},
	\end{align}
	where a typical binding agent $Y$ is graphite (\ce{C6}) \cite{Wang2012}. The electrolyte is composed of a lithium salt in solvent and dissociates according to,
	\begin{align}
	\ce{LiA
		<=>
		\ce{Li+} + \ce{A-},\label{eqn:chem}
	}
	\end{align}
	where typical examples of the anion $A$ are \ce{PF6}, \ce{AsF6}, \ce{ClO4}, and \ce{BF4} \cite{Wang2012}.
	
	The separator is a perforated micro-plastic so as to be electrically insulated yet allow for the flow of ions between the electodes.  This separator is necessary to prevent the two electrodes from touching and causing a short circuit, which would negatively impact battery performance and potentially cause safety issues such as an explosion, of which many incidents have been reported \cite{Abada2016}. A separator volume element can be decomposed into two subdomains corresponding to inactive solid material ($\Omega_\text{ia}$) and void space filled with electrolyte ($\Omega_\text{e}$).
	
	Following the pioneering work of Newman \etal\cite{Johnson1971, Newman2004, Newman1975, Newman1962}, we will model the solid and liquid phase in the three cell components using equations for conservation of mass and charge and techniques from volume averaging \cite{Bear1972, Biesheuvel2010, Fowler1997, Kaviany2012, Whitaker1969}. Volume averaging is also used in models of deionization processes where current is applied to porous electrodes in aqueous solutions \cite{Biesheuvel2010,Biesheuvel2011,Mirzadeh2014}.
	
	The volume-average approach to modelling lithium transport in the solid phase of the electrode can be related to the pseudo-two-dimensional (P2D) approach developed by Doyle \etal\cite{Doyle1993}.  In the P2D model, the solid matrix of the electrode is envisioned as a collection of spherical particles.  Accounting for solid-phase lithium transport amounts to solving the radial diffusion equation at specific points in the macroscale domain.  Volume averaging the P2D geometry leads to the model that we will consider and simplifies the problem geometry at the expense of losing information about the particle surface concentration.  If diffusion of lithium in the particles is fast, then the bulk and surface concentrations will be roughly equal and the volume-averaged concentration will be an accurate representation of the microscale composition.  One exception to this argument is the case of phase-separating electrodes, which undergo a mosaic instability whereby (de)lithiation occurs in isolated groups of particles in the electrode \cite{Li2014Nature} rather than homogeneously across the electrode.  Although resolving such features requires a P2D-style model that explicitly accounts for the microscale, volume-averaged models have been shown to produce remarkably similar predictions of macroscopic quantities such as cell potential \cite{Orvananos2014}. The reason for this will be explained below.
	
	To facilitate the asymptotic analysis and the identification of the primary electrochemical processes that occur during battery operation, we make the following modelling assumptions:
	\begin{itemize}
		\item{The lithium-ion cell is one dimensional.}
		\item{The temperature remains constant.}
		\item{The material properties are independent of composition.}
		\item{Dilute solution theory is used to describe the electrolyte.}
		\item{Phase separation in the electrodes is not explicitly considered.}
		\item{The electrostatic double layer that forms at the matrix-pore interface follows the Helmholtz model.}
	\end{itemize}
	One-dimensional geometry is motivated by the small aspect ratio ($L/H \simeq 10^{-3}$) of a typical battery cell \cite{Li2014,Safari2011}.  Although heat generation can be significant \cite{Zhao2016}, we assume there is sufficient heat exchange with the surrounding cells and the environment to maintain a constant temperature.  Val{\o}en and Reimers~\cite{Valoen2005} measured the diffusivity and ionic conductivity as a function of lithium salt concentration and showed that neither parameter changes its order of magnitude. A similar conclusion is reached from the data of Sethurajan \etal\cite{Sethurajan2015}. Along with these observations, the change in electrolyte composition is expected to be small (as verified below), motivating the use of dilute solution theory and constant parameters.
	
	Phase separation occurs in many electrode materials (e.g., \ce{LiFePO_4} or LFP).  In addition to triggering the mosaic instability described above, phase separation also plays a key role in controlling the open-circuit potential \cite{Ferguson2012, Ferguson2014}.  While neglecting phase separation may seem like a severe limitation of our proposed model, it is possible to partially account for the electrochemical impact of this process through the use of an appropriate chemical potential \cite{Orvananos2014, Thomas2017} (or equivalently an open-circuit potential) in the reaction kinetics. In \cref{sec:data} we will show that the asymptotically reduced model derived here can accurately predict the experimental discharge curves of an LFP cell using an empirical open-circuit potential.
	
	When the charged solid matrix contacts a liquid electrolyte, an electrostatic double layer forms as co- and counter-charges are repelled and attracted respectively. This layer typically has two parts: a Stern layer where counterions adhere to the matrix surface, creating a molecular dielectric with a fixed capacitance; and a diffuse layer where charges are free to move, diminishing the electrostatic effects of the charged matrix with distance. The Helmholtz model assumes that the Stern layer is much thicker than the diffuse layer, ignoring the latter's effect on the potential difference between the solid and electrolyte. This assumption is frequently used in lithium-ion battery modelling \cite{Li2014,Safari2011,Safari2011b,Smith2017}, although inclusion of diffuse-layer effects has been considerably discussed in models of deionization \cite{Biesheuvel2011,He2018,Singh2018}. The main impact of choosing how to model the electrostatic double layer comes in the form of reaction kinetics as we will discuss in \cref{sec:reaction_kinetics}.
	
	\subsection{Bulk equations}\label{sec:volumeaverage}
	The roman subscript $\text{i} = \text{n}$, $\text{p}$, $\text{s}$ is used to denote the negative electrode, positive electrode, and separator, respectively. The notation $\mn{\psi}{j}{i}$ therefore represents the quantity $\psi_j$ in component $\text{i}$. Due to the abundance of literature based on porous electrode theory, we will present our model in non-dimensional form. However, the full dimensional equations and their derivation appears in \cref{sec:fullmod} for posterity. 
	
	In non-dimensionalising, space is scaled with the length $L$ of the cell, time with the diffusive time scale of lithium in the electrolyte $L^2/D_L$, and current densities with the nominal applied current density $i_0$. Concentrations and electric potentials (including open-circuit potentials) are written as the deviation from their initial values and scaled with the change due to electrochemical reactions $(i_0 L)/(F D_L)$ and the thermal voltage $R T_a/F$, respectively, where $R$ is the universal gas constant, $F$ is Faraday's constant and $T_a$ is the ambient temperature.
	
	Volume averaging for conservation of mass and charge of the active solid phase in electrode $\text{i}$ results in
	\subeq{
		\label{sys:electrode_solid}
		\begin{align}
		\pderiv{\csi}{t}&=\Di\pderiv[2]{\csi}{x} +\pderiv{\isi}{x}, \label{eqn:nd_csi}\\
		\mn{\nu}{a}{i} \isi&=-\pderiv{\Phisi}{x},\label{eqn:ohma}\\
		\phisi\pderiv{\isi}{x} &=-\Gi \left(\gibar +\Ci \pderiv{}{t}(\Phisi-\Phiei)\right),\label{eqn:nd-isi}
		\end{align}
	}where $t$ is time, $x$ is the horizontal coordinate, $\csi$ is the concentration of intercalated lithium, $\isi$ is the current density in the active solid phase, and $\Phisi$ and $\Phiei$ are the electric potential in the active solid and electrolyte, respectively. The concentration $\csi$ can be interpreted as that which arises from volume averaging over several particles in the P2D model. The parameter $\Di$ is the ratio of solid-phase to liquid-phase lithium diffusivity. The quantity $\gibar$ is the non-dimensional surface-averaged electrochemical current that is produced at the electrode-electrolyte interface, which will be defined in \cref{sec:reaction_kinetics}. Associated with it is the parameter $\Gi$ which is the ratio of current produced by surface reactions to the input current of the system. The parameter $\phisi$ is the volume fraction of active solid material, $\Ci$ is the non-dimensional capacitance associated with a double-charging layer, and $\mn{\nu}{a}{i}$ is a non-dimensional resistivity.

	Similar equations follow for the fluid phase by averaging over the electrolyte volume:
	\begin{subequations}\label{sys:electrode_liquid}
		\begin{align}
		\pderiv{\cli}{t}&=\pderiv{}{x}\left(\pderiv{\cli}{x} +\nue^{-1}\theta\left(1+\gamma\cli\right)\pderiv{\Phiei}{x}\right)+\pderiv{\iei}{x}, \label{eqn:nd_cli}\\
		\iei&=-\left(1-\Da\right)\pderiv{\cli}{x} -\nue^{-1}(1+\gamma\cli)\pderiv{\Phiei}{x},\label{eqn:ohme}\\
		\phiei\pderiv{\iei}{x}&=\Gi\left(\gibar+\Ci\pderiv{}{t}(\Phisi-\Phiei)\right).\label{eqn:nd-iei}
		\end{align}
	\end{subequations}
	Here, $\cli$ is the concentration of lithium ions in the electrolyte, $\iei$ is the electrolytic current, and $\phiei$ is the volume fraction of electrolyte. We do not explicitly model the concentration of anions because electroneutrality requires that it be the same as the lithium concentration. The parameter $\Da$ is the ratio of anion diffusivity to lithium-ion diffusivity in the electrolyte, $\theta$ is the transference number, $\nue$ is a non-dimensional electrolyte resistivity, and $\gamma$ is the relative change in lithium ion concentration from its initial value. A phase-averaged conservation of charge emerges by adding \cref{eqn:nd-isi} and \cref{eqn:nd-iei},
	\begin{align}
	\pderiv{}{x}(\phisi \isi+\phiei\iei)=0,\label{eqn:phaseneutral}
	\end{align}
	which will be used in place of \cref{eqn:nd-iei}.
	
	Finally, the non-dimensional model in the separator is
	\begin{subequations}\label{sys:separator2}
		\begin{align}
		\pderiv{\cls}{t}&=\pderiv{}{x}\left(\pderiv{ \cls}{x}+\nu_e^{-1}\theta\left(1+\gamma\cls\right)\pderiv{\Phies}{x}\right), \label{eqn:nd_cls}\\
		\ies&=-\left(1-\Da\right)\pderiv{ \cls}{x}-\nue^{-1}(1+\gamma\cls)\pderiv{\Phies}{x},\label{eqn:ohmes}\\
		\pderiv{\ies}{x}&=0,\label{eqn:ohmas}
		\end{align}
	\end{subequations}
	where the main difference to the electrolyte problem in the electrode is the absence of surface reaction currents. We can eliminate the electrolyte potential from the liquid equations in the electrode (by manipulating \cref{eqn:nd_cli} and \cref{eqn:ohme}) and separator (by manipulating \cref{eqn:nd_cls} and \cref{eqn:ohmes})  resulting in
	\subeq{
		\begin{align}
		\pderiv{\cli}{t}&=\R \pderiv[2]{\cli}{x} + (1 - \theta)\pderiv{\iei}{x}, \label{eqn:nd_cli2}\\
		\pderiv{\cls}{t}&=\R \pderiv[2]{\cls}{x}, \label{eqn:nd_cls2}
		\end{align}
	}
	where $\R = 1 - \theta (1 - \Da)$.
	
	The cell voltage, $\Delta V$, is determined as the difference in the potentials in the solid phase of the positive electrode at $x=0$ and negative electrode at $x=L$,
	\begin{align}
	\Delta V=\Phisp(0,t)-\Phisn(L,t) + \log \Up - \log \Un,\label{eqn:cellvolt}
	\end{align}
	where $(R T_a / F)\log \Ui$ represents the initial value of the dimensional open-circuit potential.  An expression for $\Ui$ in terms of reaction constants is given in \cref{eqn:UVdef}.
	
	\subsection{Reaction kinetics}\label{sec:reaction_kinetics}
	The surface reaction currents $\gibar$ are described by the Butler--Volmer kinetics \cite{Doyle1993, Fuller1994, Newman2004, Newman1975} in the Helmholtz limit valid for thin electric double layers at high salt concentrations \cite{Smith2017}. This means that we will neglect the Frumkin correction which explicitly creates a dependence of the reaction rate on the local electric field near the surface of the solid matrix (see \cref{sec:kinetics} for more details). The Helmholtz assumption could limit practically achievable power densities at high discharge rates when the electrolyte salt becomes depleted \cite{Smith2017}.
	
	Along with the Helmholtz assumption, 
	we prescribe a theoretical open-circuit potential following Newman \cite[page 212]{Newman2004} (see \cref{sec:kinetics}). This defines the non-dimensional Butler--Volmer reaction kinetics as
	\begin{subequations}\label{sys:nondimbutler}
		\begin{align}
		\gibar&=\ji\left(\exp\left[(1-\betai)\etai\right]-\exp\left[-\betai\etai\right]\right),\\
		\ji&=\left(1+\deltai\gamma\csi\right)^{\betai}\left(1-\deltai\xii(1-\xii)^{-1}\gamma\csi\right)^{1-\betai}(1+\gamma\cli)^{1-\betai}. \label{eqn:nd_j0i}
		\end{align}
	\end{subequations}
	The surface overpotential is defined by $\etai=\Phisi-\Phiei-\mnn{U}{i}$, with
	\begin{align}
	\mnn{U}{i} = \log \Vi,
	\qquad
	\Vi=\frac{(1+\gamma\cli)[1-\deltai\xii(1-\xii)^{-1}\gamma\csi]}{1+\deltai\gamma\csi}, \label{eqn:butleretadef}
	\end{align}
	denoting the open-circuit potential. The parameter $\betai$ is a symmetry factor, $\deltai$ is the ratio of initial lithium in electrolyte to solid, and $\xii$ is the ratio of the initial solid concentration to the maximum amount allowed in the electrode. This is also the initial state of charge. By extending the form of the open-circuit potential \cref{eqn:butleretadef}, it is possible to account for additional physics such as phase separation \cite{Ferguson2012, Ferguson2014, Orvananos2014} and multiple lithiation stages \cite{Ferguson2014, Thomas2017} in the volume-averaged model.
	
	\subsection{Boundary and initial conditions}
	The electrolyte is free to flow between the voids of the electrodes and separator. Therefore, we require the concentration and molar flux of lithium ions and the current density in the electrolyte, as well as the electrolyte potential, to be continuous. Continuity of flux and current can be simplified to yield continuity in the derivatives of lithium concentration and electrolyte potential:
	\begin{subequations}
		\label{bc:se_fluid}
		\begin{alignat}{2}
		\cli - \cls &= 0, \quad &&x = x_p,\,x_n;\\
		\phiei\pderiv{\cli}{x}-\phies\pderiv{\cls}{x}&=0,\quad &&x = x_p,\,x_n;\label{bc:dcli_cont}\\
		\phiei\pderiv{\Phiei}{x}-\phies\pderiv{\Phies}{x}&=0,\quad &&x = x_p,\,x_n;\label{bc:dPhiei_cont}\\
		\Phiei-\Phies &= 0,\quad &&x = x_p,\,x_n. \label{bc:phi_e_cont}
		\end{alignat}
	\end{subequations}
	The volume fractions appearing in \cref{bc:se_fluid} account for differences in the porosity of each material and arise from the process of averaging the microscopic boundary conditions. The solid component of the separator is electrically inactive and therefore no current can pass through it. 
	\begin{alignat}{2}
	\isi&=0, &\quad  x&=x_p,\,x_n.\label{bc:noisi}
	\end{alignat}
	
	The electrode surfaces at $x = 0$ and $x = 1$ are in contact with current collectors which enable electric charge to be injected into and extracted from the cell during charging and discharging. We focus on the case of battery discharging and therefore assume that a non-dimensional current density of $\mathcal{I}$ is being drawn from the positive electrode. This value is also known as the C-rate of the battery, where $\mathcal{I}=1$ is equivalent to the battery fully discharging over an hour at its nominal rate (see \cref{sec:BC} for more details). The discharge boundary condition is
	\begin{align}
	\phisp\isp=-\mathcal{I}, \quad x = 0, \label{bc:iapp}
	\end{align}
	where the negative sign on the right-hand side indicates a discharge process.
	
	Without loss of generality, we can set the electrolyte potential in the negative electrode to zero at the electrode-collector interface, leading to
	\begin{align}
	\Phien=0, \quad x=1.\label{bc:phi_e_ground}
	\end{align}
	The current collectors are impermeable and therefore the molar fluxes, and hence the current, of the electrolyte must vanish at the electrode-collector interfaces,
	\begin{align}
	\iei = 0, \quad x =0,\, 1.\label{bc:iei}
	\end{align} 
	Similar to the electrode-separator interfaces, we can combine the vanishing molar flux condition for both lithium and anionic species which produces a Neumann condition for the lithium concentration,
	\begin{align}
	\pderiv{\cli}{x} = 0, \quad x = 0,\, 1.\label{bc:dcli}
	\end{align}
	The mass flux of the solid lithium must also vanish at the electrode boundaries:
	\begin{align}
	\pderiv{\csi}{x} = 0, \quad x=0,\,x_p,\,x_n,\,1,\label{bc:dcsi}
	\end{align}
	The initial conditions are given by $\csi(x,0) = 0$, $\cli(x,0) = 0$, $\Phiei(x,0) = 0$, and $\Phisi(x,0) = 0$ because of the choice of non-dimensionalisation.
	
	\section{Asymptotic reduction} \label{sec:asymptotics}
	
	The main result of this paper is to systematically derive a reduced model for an LIB of the form:
	\begin{subequations}\label{sys:compred}
		\begin{alignat}{2}
		\Cn\deriv{\Phisn}{t}&=\frac{\mathcal{I}}{\Gn(1-x_n)}-\gnbar(\Phisn,\csn),&\qquad \csn=-\frac{\mathcal{I}}{\phisn(1-x_n)}t;\\
		\Cp\deriv{\Phisp}{t}&=-\frac{\mathcal{I}}{\Gp x_p}-\gpbar(\Phisp, \csp),&\qquad \csp=-\frac{\mathcal{I}}{\phisp x_p}t,
		\end{alignat}
	\end{subequations}
	with $\Phisi(0) = 0$. To obtain \cref{sys:compred}, we carry out a preliminary reduction of the full non-dimensional model using regular perturbation theory, followed by a detailed asymptotic analysis using singular perturbation theory. This is now discussed in detail.
	
	\subsection{Preliminary model reduction}

	Physical constants for different batteries are presented throughout the literature \cite{Amiribavandpour2015, Li2014, Ong1999, Richardson2012} and generally result in all of the parameters in \cref{sys:electrode_solid}, \cref{sys:electrode_liquid}, and \cref{sys:separator2} being small except for $\Da, \Gi, \xii, \deltai$, and $\Ui$.  The order one assumptions for $\deltai$ and $\xii$ are generally only true for the initially lithiated electrode.  
	
	Using the parameter sizes considered above, the model is naturally reduced by neglecting all parameters which are less than $\Ord{1}$ in size. We can set $\Di \to 0$ in \cref{eqn:nd_csi} as the no-flux conditions for $\csi$ given by \cref{bc:dcsi} are consistent with the spatially uniform initial condition so boundary layers are avoided. Taking $\nue \to 0$ in \cref{eqn:ohme} and \cref{eqn:ohmes} shows that the electrolyte potential $\Phiei$ is constant in space and through the continuity and grounding conditions \cref{bc:phi_e_cont} and \cref{bc:phi_e_ground} must be zero everywhere, $\Phiei \equiv 0$. Similarly, taking $\mn{\nu}{a}{i} \to 0$ in \cref{eqn:ohma} shows that the active solid potential is constant in space. Finally, although $\Ci$ and $\gamma$ are small, setting them to zero leads to singular limits representing distinguished time regimes which we study using matched asymptotic expansions. The singular limit for $\Ci$ arises because it is multiplying a time derivative of the potential marking an early time regime where capacitance effects are relevant.  The singular limit for $\gamma$ is less obvious but arises from the reaction currents in \cref{eqn:nd_j0i} and $\Vi$ in \cref{eqn:butleretadef}, which suggest the possibility of a regime where the concentrations are $\Ord{\gamma^{-1}}$ in size, corresponding to the depletion/saturation of lithium in the electrodes.
	
	Taking the limit as $\Di \to 0$, $\nue \to 0$, and $\mn{\nu}{a}{i} \to 0$, while retaining the parameters $\Ci$ and $\gamma$, leads to a simplified set of bulk equations given by
	\begin{subequations}\label{sys:reduceelec}
		\begin{align}
		\pderiv{\csi}{t}&=\pderiv{\isi}{x},\label{eqn:csi}\\
		\phisi\pderiv{\isi}{x} &=-\Gi \left(\gibar +\Ci \deriv{\Phisi}{t}\right),\label{eqn:isi}\\
		\pderiv{\cli}{t}&= \R \pderiv[2]{\cli}{x} + (1 - \theta) \pderiv{\iei}{x}, \label{eqn:cli}
		\end{align}
	\end{subequations}
	for the electrodes and
	\begin{subequations}\label{sys:reduces}
		\begin{align}
		\pderiv{\cls}{t}&= \R \pderiv[2]{\cls}{x} \label{eqn:cls}
		\end{align}
	\end{subequations}
	for the separator. Governing equations for the electrolyte current are not required as the one-dimensional charge conservation condition \cref{eqn:phaseneutral} can be integrated to find that
	\begin{align}
	\phisi\isi+\phiei\iei=-\mathcal{I} \label{eqn:globalneutral1D}
	\end{align}
	in each of the cell components, where the boundary conditions \cref{bc:noisi}, \cref{bc:iapp}, and \cref{bc:iei} have been used. The Butler-Volmer kinetics are given by \cref{sys:nondimbutler} and \cref{eqn:butleretadef} with a reduced overpotential $\etai = \Phisi - \log(\Vi)$.
	
	The boundary conditions for this simplified model are given by
	\subeq{
		\label{eqn:reduceBC}
		\begin{alignat}{2}
		\phisp\isp&=-\mathcal{I}, \quad &&x = 0; \label{red_bc:isp1}\\ 
		\cli - \cls &= 0, \quad &&x = x_p,\,x_n;\\
		\phiei\pderiv{\cli}{x}-\phies\pderiv{\cls}{x}&=0, \quad && x=x_p,\, x_n;\\
		\isi&=0,\quad &&x = x_p,\,x_n; \label{red_bc:isi}\\
		\pderiv{\cli}{x}&=0,\qquad && x=0,\,1,
		\end{alignat}
	}
	while the initial conditions are $\csi=\cli=\Phisi=0$.

	We now proceed to solve the simplified model using asymptotic methods. Our approach exploits the fact that, based on singular limits for $\Ci$ and $\gamma$, there are three key regimes that occur during battery discharge. First, there is a small-time regime, given by $t = \Ord{\Ci}$, that captures the rapid formation of double charging layers at the electrode-electrolyte interfaces due to the instantaneous application of current to the cell. In the first regime, capacitance effects play a key role and composition changes are negligible. In the second time regime, defined by $t = \Ord{1}$, capacitance effects become negligible. Electrochemical reactions lead to $\Ord{1}$ changes in the concentration of intercalated lithium ions in the electrodes and diffusive transport begins in the electrolyte. In the third and final regime, given by $t = \Ord{\gamma^{-1}}$, the electrodes become fully saturated and depleted of lithium, corresponding to a drained battery. The first two regimes have previously been identified in models of desalination with the first regime termed the super-capacitive regime where charge storage occurs while the second is the capacitive dionization regime where salt is removed \cite{Biesheuvel2010,Biesheuvel2011}.

	\subsection{First regime: double charging layer}\label{sec:ass1}
	
	The first regime is captured by choosing a time scale that balances both terms on the right-hand side of \cref{eqn:isi}. Typically \cite{Li2014, Safari2011}, the material properties are such that $\Cn\ll\Cp$, leading to two sub-regimes that must be considered. Thus, we first calculate solutions for $t = \Ord{\Cn}$ and then focus on the case when $t = \Ord{\Cp}$. 
	
	In the first subregime, we let $t = \Cn \tilde{t}$ in \cref{sys:reduceelec} and \cref{sys:reduces}. Upon taking $\Cn \to 0$ and $\gamma \to 0$ with $\csi = \Ord{1}$ and $\cli = \Ord{1}$, we obtain
	\begin{align}
	\pderiv{\cli}{\tilde{t}}=\pderiv{\csi}{\tilde{t}} = \deriv{\Phisp}{\tilde{t}} = 0.\label{eqn:capconc}
	\end{align}
	Thus, the concentrations remain unchanged from their initial value: $\cli \equiv 0$ and $\csi \equiv 0$. The solid potential in the positive electrode is $\Phisp \equiv 0$.  For the negative electrode, we find
	\begin{align}
	\phisn\pderiv{\isn}{x}=-\Gn\left(\gnbar+\deriv{\Phisn}{\tilde{t}}\right).
	\label{eqn:R11_psisn}
	\end{align}
	Since the solid- and liquid-phase concentrations of lithium remain at zero then $\Vi=1$ and $\etai=\Phisi$. Therefore, $\gnbar$ is now solely a function of time and so \cref{eqn:R11_psisn} can be integrated in space using \cref{red_bc:isp1} and \cref{red_bc:isi} to yield a differential equation for $\Phisn$ given by
	\begin{align}
	\deriv{\Phisn}{\tilde{t}} = \frac{\mathcal{I}}{\Gn(1 - x_n)} - \left[\exp((1-\betan)\Phisn)-\exp(-\betan\Phisn)\right], \label{eqn:cap1DE}
	\end{align}
	where $\Phisn(0) = 0$.  Using the initial condition, we see that $\mathrm{d} \Phisn / \mathrm{d} \tilde{t} > 0$ when $\tilde{t} = 0$. Thus, the potential in the negative electrode will increase in time until it reaches a steady state $\Phisn^*$ given by
	\begin{align}
	\exp((1-\betan)\Phisn^*)-\exp(-\betan\Phisn^*) = \frac{\mathcal{I}}{\Gn(1 - x_n)}. \label{eqn:etan_star}
	\end{align}
	When $\betan = 1/2$, which is often considered in other models and corresponds to symmetric anodic and cathodic reactions, an implicit solution to \cref{eqn:cap1DE} can be obtained (see \cref{sec:etasol}). Using \cref{eqn:cellvolt}, the cell potential in this sub-regime, $\Delta V_\mathrm{I}^\mathrm{n}$, is
	\begin{align}
	\Delta V_{\rm I}^\mathrm{n} =\log \Up -\Phisn - \log \Un,\label{eqn:VI}
	\end{align}
	with $\Phisn$ computed from \cref{eqn:cap1DE}.
	
	The next capacitance sub-regime can by analysed by letting $t=\Cp\check{t}$ and taking $\Cp \to 0$ and $\gamma \to 0$ with $\csi = \Ord{1}$ and $\cli = \Ord{1}$, which still leaves the concentrations unchanged and results in the electrode kinetics
	\begin{subequations}
		\begin{align}
		0 &= \frac{\mathcal{I}}{\Gn(1 - x_n)} - \left[\exp((1-\betan)\Phisn)-\exp(-\betan\Phisn)\right], \label{eqn:cap2gn} \\
		\deriv{\Phisp}{\check{t}} &= -\frac{\mathcal{I}}{\Gp x_p} - \left[\exp((1-\betap)\Phisp)-\exp(-\betap\Phisp)\right], \label{eqn:cap2gp}
		\end{align}
	\end{subequations}
	which have come from integrating \cref{eqn:isi} as once again $\gibar$ is space independent.  Equation \cref{eqn:cap2gn} prescribes a steady potential in the negative electrode phase, $\Phisn(\check{t}) \equiv \Phisn^*$, which matches to that in the previous sub-regime.  The initial condition for \cref{eqn:cap2gp} is also obtained by matching to the solution in the previous sub-regime, which yields $\Phisp(0) = 0$. Equation \cref{eqn:cap2gp} describes a decreasing potential in the positive electrode to a steady state given by
	\begin{align}
	\exp((1-\betap)\Phisp^*)-\exp(-\betap\Phisp^*) = -\frac{\mathcal{I}}{\Gp x_p}. \label{eqn:etap_star}
	\end{align}
	As before, analytical solutions for $\Phisp$ and $\Phisp^*$ can be obtained when $\betap = 1/2$. The cell potential in this region, $\Delta V_{\rm I}^\mathrm{p}$, is
	\begin{align}
	\Delta V_{\rm I}^{\rm{p}} =\Phisp + \log \Up-\Phisn^* - \log \Un,\label{eqn:VII}
	\end{align}
	where $\Phisn^*$ is given by \cref{eqn:etan_star} and $\Phisp$ is determined by \cref{eqn:cap2gp}.
	
	\subsection{Second regime: diffusion in liquid}\label{sec:ass2}
	We now move on to the second regime where $t=\Ord{1}$.  Matching to the solutions in the first regime implies that the concentrations $\csi$ and $\cli$ must be $\Ord{1}$ in magnitude. Thus, we can take $\gamma \to 0$ to show that $\Vi=1$  and $\gibar$ remains independent of space. Equation \cref{eqn:isi} can be integrated as in \cref{sec:ass1} and the limits $\Ci \to 0$ can be taken to obtain
	\begin{align}
	\Phisi(t) \equiv \Phisi^*,
	\label{eqn:gequal}
	\end{align}
	which automatically matches to the solutions for the overpotential in the first regime. The cell voltage in this region, $\Delta V_{\rm II}$,  is given by
	\begin{align}
	\Delta V_{\rm II}=\Phisp^* + \log \Up - \Phisn^* - \log \Un \label{eqn:VIII}
	\end{align}
	and is constant in time. We also have that the active solid current densities are given by
	\begin{subequations}\label{sys:currents}
		\begin{align}
		\isn=-\frac{\mathcal{I}}{\phisn(1-x_n)}(x-x_n),\qquad
		\isp=-\frac{\mathcal{I}}{\phisp x_p}(x_p-x),
		\end{align}
	\end{subequations}
	which we can substitute into \cref{eqn:csi} for each of the electrodes to find that the intercalated lithium-ion concentrations are
	\begin{align}
	\csn=\cnc t, \qquad
	\csp=\cpc t, \label{sys:t1csi}
	\end{align}
	where we have used the matching conditions $\csi \sim 0$ as $t \sim 0$. 
	In principle, the concentration of lithium ions in the electrolyte, $\cli$, can be obtained using separation of variables as has been utilised in models without intercalation kinetics \cite{Doyle1997, Guduru2012}. However, for our purposes, it is sufficient to consider the steady-state concentration profile given by
	
	\subeq{
		\begin{align}
		c_L^*=\fshat{2}
		\begin{cases}
		\left(x^2 / x_p+(2 \phie / \phies - 1) x_p+\mathcal{B}\right),&0\leq x \leq x_p,\\
		\left((2\phie/\phies) x+\mathcal{B}\right),&x_p\leq x \leq x_n,\\
		\left(1+(2 \phie / \phies - 1)x_n-(1-x)^2/(1-x_n)+\mathcal{B}\right),& x_n\leq x\leq 1,
		\end{cases} 
		\label{eqn:clanal}
		\end{align}
		where 
		\begin{align}
		\begin{split}
		\mathcal{B} = &\frac{\phie}{\phie x_p + \phies(x_n - x_p) + \phie (1 - x_n)}\times \\
		&\left\{
		\frac{1}{3}\left[(1 - x_n)^2 - x_p^2\right] + 2 \left(1 - \frac{\phie}{\phies}\right)\left[x_n(1-x_n)+x_p^2\right]
		-1
		\right\}.
		\end{split}
		\end{align}
	}
	In deriving \cref{eqn:clanal}, we have used the fact that
	\begin{align}
	\int_{0}^{x_p} \phie \clp\,\diff x + \int_{x_p}^{x_n} \phies \cls\,\diff x + \int_{x_n}^{1} \phie \cln\,\diff x= 0
	\end{align}
	for all time, which arises from the no-flux boundary conditions at the electrode-collector interfaces and continuity of flux across the electrode-separator interfaces, implying that the total concentration of lithum in the electrolyte is a conserved quantity. 
	
	\subsection{Third regime: electrode saturation/depletion}\label{sec:ass3}
	The linear growth and decay of the concentration of intercalated lithium in \cref{sys:t1csi} necessitates a large-time regime where the finite capacity of the electrodes must be taken into consideration. Mathematically, this means capturing the composition dependence of the Butler--Volmer kinetics \cref{sys:nondimbutler}. In the first and second regimes, this dependence could be removed by taking the limit as $\gamma \to 0$ with $\csi = \Ord{1}$. We now account for large changes in $\csi$ which alter the details of this limit.
	
	An examination of the expression for $\gibar$ given by \cref{sys:nondimbutler} shows that the composition dependence becomes relevant when the concentrations $\csi$ become $\Ord{\gamma^{-1}}$ in size. From \cref{sys:t1csi}, this concentration scale corresponds to a time scale of $t=\Ord{\gamma^{-1}}$. Thus, in the third regime, we write $t = \gamma^{-1} \hat{t}$, $\csi = \gamma^{-1} \csihat$, and $\Phi = \Phisihat$. There is no need to rescale the concentration of lithium in the electrolyte since matching to the second regime implies $\cli = \Ord{1}$. With this scaling, it is then possible to take $\Ci \to 0$ and $\gamma \to 0$ as before. The limit $\gamma \to 0$ removes the dependence of $\gibar$ on $\cli$, however the dependence on $\csihat$ is retained because $\Vi\neq1$. The matching conditions for $\csihat$ are given by
	\begin{align}
	\csnhat = \cnc \hat{t}, \qquad
	\csphat = \cpc \hat{t}, \label{sys:csihat}
	\end{align}
	as $\hat{t} \sim 0$, which imply that $\csihat$ and hence $\gibar$ will be independent of space for all time. The same procedure as in the first and second time regimes can then be used to obtain solutions in the third regime. The concentrations of intercalated lithium are given by \cref{sys:csihat}, the concentration of lithium ions in the electrolyte is constant in time and given by \cref{eqn:clanal}, and the current densities are those in \cref{sys:currents}. The electrode kinetics, can be written in terms of the concentrations as
	\subeq{
		\label{eqn:lategbar}
		\begin{align}
		\frac{\mathcal{I}}{\Gn(1 - x_n)} &= (1+\deltan\csnhat)\exp((1-\betan)\Phisnhat) \nonumber \\
		&\quad-(1-\deltan\xin(1-\xin)^{-1}\csnhat)\exp(-\betan\Phisnhat), \\
		-\frac{\mathcal{I}}{\Gp x_p} &= (1+\deltap\csphat)\exp((1-\betap)\Phisphat) \nonumber \\
		&\quad-(1-\deltap\xin(1-\xip)^{-1}\csphat)\exp(-\betap\Phisphat).
		\end{align}
	}
	The cell voltage in this regime, $\Delta V_{\rm III}$, is given by
	\begin{align}
	\Delta V_{\rm III}=\Phisphat + \log \Up - \Phisnhat - \log \Un,\label{eqn:VIV}
	\end{align}
	where $\Phisihat$ comes from solving \cref{eqn:lategbar} with the time-dependent concentrations given by \cref{sys:csihat}.  
	
	An examination of \cref{eqn:lategbar} reveals that the electrode potential $\Phisihat$ becomes singular at finite concentrations given by
	\begin{subequations}\label{sys:Csilims}
		\begin{alignat}{2}
		\csihat &= -\frac{1}{\deltai}, &\quad \text{leading to }\Phisihat &\to\infty,\label{eqn:poslim}\\
		\csihat &= \frac{1-\xii}{\deltai\xii}, &\quad \text{leading to }\Phisihat &\to-\infty.\label{eqn:neglim}
		\end{alignat}
	\end{subequations}
	These are precisely the non-dimensional variants of the two limiting (dimensional) concentrations, $\csi = 0$ and $\csi = \csimax$, respectively. 
	
	At first appearance, it seems the physically infeasible unbounded growth and decay of the concentration of intercalated lithium has not been resolved as the solutions \cref{sys:csihat} indicate that the linear dependence on time persists.  However, in consideration of the limits in \cref{sys:Csilims}, finite-time blow-up occurs in the electric potential as these terminal concentrations are approached.  In the negative electrode, $\csnhat$ decreases so \cref{eqn:poslim} gives the terminal value of $\csnhat$. Similarly, \cref{eqn:neglim} gives the limiting value for $\csphat$. Using \eqref{sys:csihat}, the terminal concentrations in \cref{sys:Csilims} correspond to blow-up times given by
	\begin{align}
	\hat{t}_n=\frac{\phisn(1-x_n)}{\deltan\mathcal{I}},\qquad \hat{t}_p=\frac{\phisp x_p(1-\xip)}{\deltap\xip\mathcal{I}}.\label{eqn:tblow}
	\end{align}
	Physically, the finite-time blow-up corresponds to a failure of the model where a constant discharge/charge current $\mathcal{I}$ is no longer feasible. The battery stops operating at
	\begin{align}
	\hat{t}_c=\min\{\hat{t}_n,\hat{t}_p\}.\label{eqn:Tc}
	\end{align}
	Finite-time blow-up is rarely mentioned in other models as simulations are typically terminated based on a threshold value of the cell potential \cite{Li2014}.
	
	\subsection{Construction of the composite reduced model}
	
	We can now construct the composite model \cref{sys:compred}, and generalise it to account for general open-circuit potentials, by recognizing that regardless of the open-circuit potential, the electrolyte concentration always reaches an $\Ord{1}$ steady state, $c_L^*$ given by \cref{eqn:clanal}, in $\Ord{1}$ time as per the analysis of regime 2 in \cref{sec:ass2}. Consequently, we can ignore the electrolyte concentration when solving for the electric potentials. We also have that the solid-phase concentrations $\csi$ are spatially uniform, linear functions of time given by \cref{sys:t1csi}, which are valid throughout all three regimes. Therefore, the main impact of each regime is to change the voltage dynamics, primarily from open-circuit conditions in regime 1 to saturation/depletion conditions in regime 3. For this reason we can pose the composite reduced model by retaining the two singular contributions, leading to the model given by \cref{sys:compred}, where $\gibar$ is the full concentration-dependent reaction kinetics defined by \cref{sys:nondimbutler}, but now with an arbitrary form of the open-circuit potential $\mnn{U}{i}$. For consistency with the non-dimensionalisation, the initial value of $\mnn{U}{i}$ must be zero, with the initial value of the dimensional open-circuit potential being equal to $(R T_a / F) \log \Ui$. The composite model \cref{sys:compred} has both the powerful simplicity of the asymptotic reduction and the versatility to easily handle a variety of open-circuit potentials $\mnn{U}{i}$ that extend those given by \cref{eqn:butleretadef}
	
	\section{Comparison with numerics}\label{sec:comparison}
	
	We now compare the asymptotic reduction from \cref{sec:asymptotics} to simulations of the full model to assess the accuracy of our approach. We take $\Da=\Gn=\Gp=\mathcal{I}=\R=1$ as these have been assumed to be $\Ord{1}$ in size.  We also take $\betai = 1/2$ assuming symmetry in the anodic and cathodic current. For the small parameters, we take $\Dn=\Dp=\gamma=\mnn{\nu}{s}=\mnn{\nu}{e}=\Cn=10^{-2}$. We also take $\Up=\Un=2$ since $\log\Ui$ appears throughout and we wish to avoid $\log\Ui=0$. We take $\Cp=0.1$ in order to satisfy $\Cn\ll\Cp\ll1$ and explicitly showcase the two capacitance sub-regimes.  For symmetry, we consider $x_p=0.34$ and $x_n=0.67$ so that each domain takes up approximately a third of the battery cell and also take $\theta=0.5$ so that the effective charge is carried equally by lithium and the electrolytic salt.  Since the porosity of electrodes is quite small, we take $\phie=\phies=0.33$ and also assume that half of the volume is occupied by active material, i.e. $\phisi=0.5$.  Due to our consideration of a discharge process, we will assume that the negative electrode is mostly saturated in lithium while the positive electrode is depleted and thus take $\xin = 0.9$ and $\xip = 0.05$. We will further assume that both electrodes have the same maximal concentration. Therefore, by definition of $\xii$ and $\deltai$, the ratio $\xip/\xin=\deltan/\deltap$ must be held constant.  As such, we fix $\deltap=1$ which restricts $\deltan=5.56\tten{-2}$. These values are roughly based on those obtained using the dimensional parameter values in \cref{sec:params}.
	
	We simulate the full model, \cref{sys:electrode_solid}, \cref{sys:electrode_liquid}, \cref{sys:separator2} using a second-order central difference discretisation in space and backward Euler discretisation in time as detailed in \cref{sec:numerics}. For all simulations, we take 50 interior cell-centres in each of the three domains and compute until the finite-time blow-up induced by \cref{eqn:tblow}.  For the chosen parameters, this occurs at the non-dimensional time $t=297$ following \cref{eqn:Tc} which corresponds to $t_n$.
	
\begin{figure}
		\centering
		\subfloat[Positive Electrode.]
		{
			\label{subfig:cspass}\includegraphics[width=0.45\textwidth]{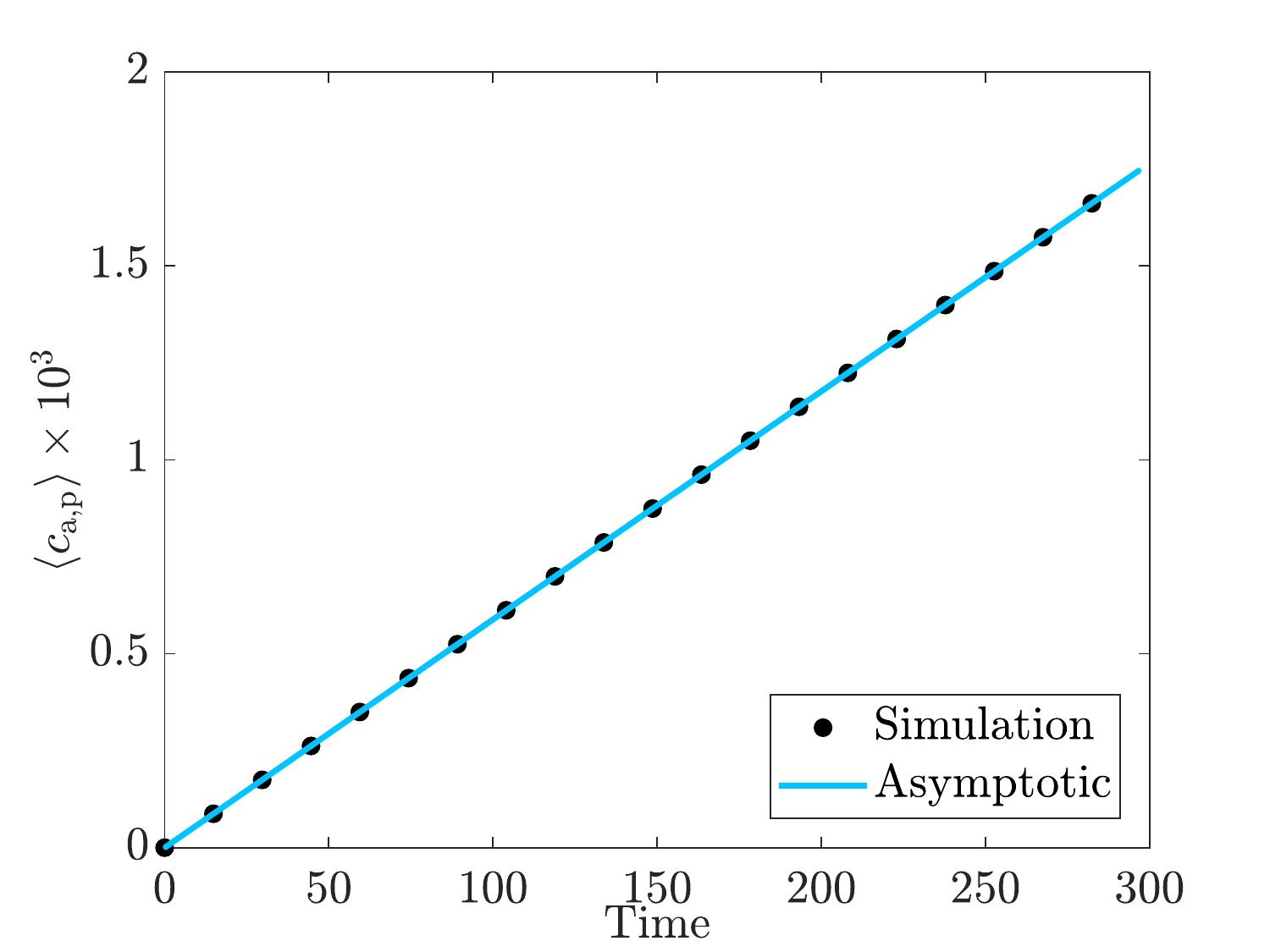}
		}
		\qquad
		\subfloat[Negative Electrode.]
		{
			\label{subfig:csnass}\includegraphics[width=0.45\textwidth]{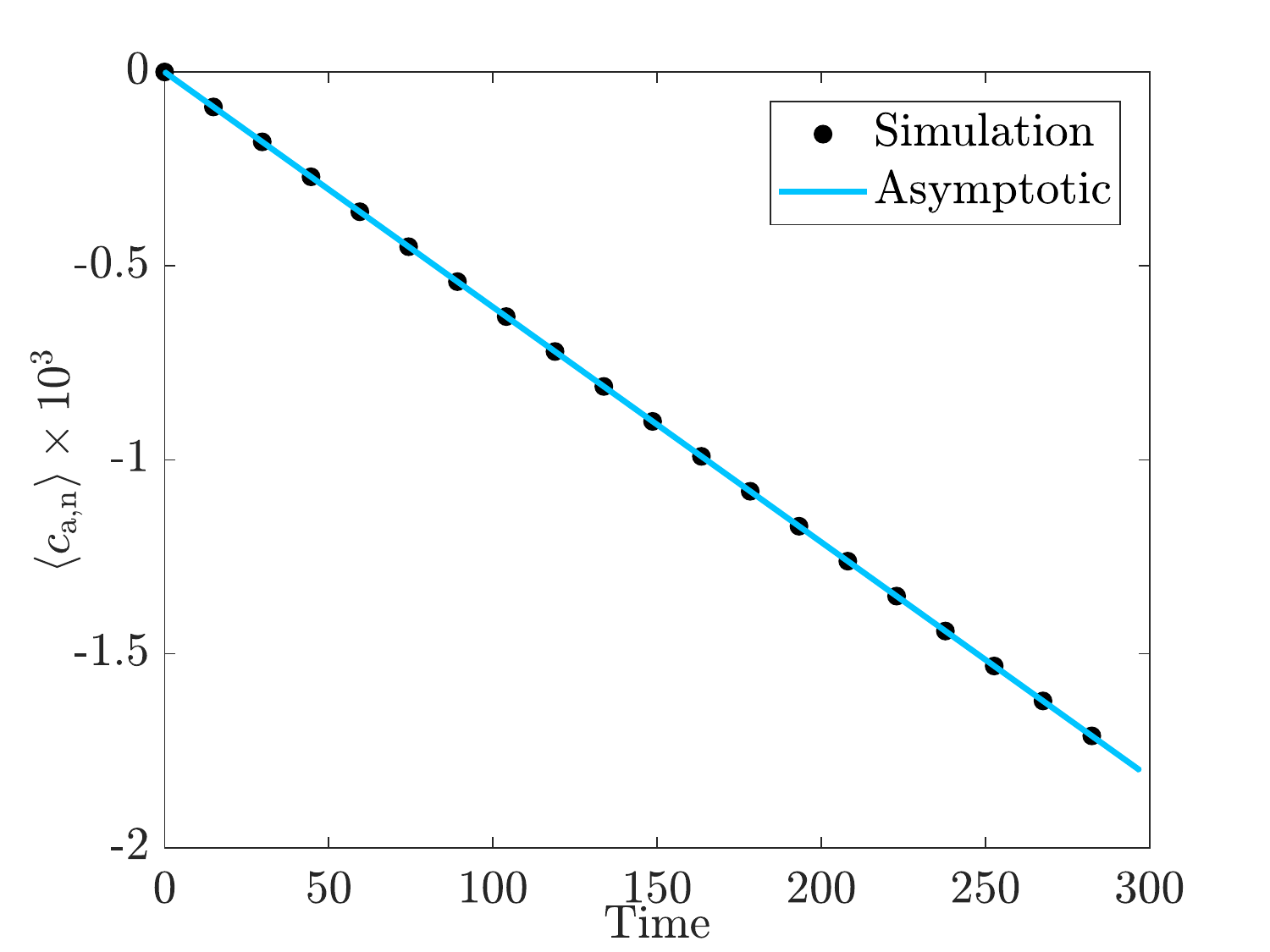}
		}
		\caption{Asymptotic solution \cref{sys:t1csi} compared to the space-averaged simulated solid lithium concentration in each of the electrodes.}
		\label{fig:csass}
	\end{figure}

	We first compare the asymptotic and numerical profiles for the solid-phase lithium concentration.  As predicted from the asymptotic analysis, the numerical profiles have weak spatial gradients.  Therefore, we take spatial averages,
	\begin{align}
	\langle \csp \rangle=\frac{1}{x_p}\int_0^{x_p}\csp(x,t)\diff x,\qquad\langle \csn \rangle=\frac{1}{1-x_n}\int_{x_n}^1\csn(x,t)\diff x,
	\end{align}
	and plot them against the asymptotic expressions for the solid-phase concentration \cref{sys:t1csi} in \cref{fig:csass}.  The agreement is excellent, with the numerical solution confirming the linear-in-time mean intercalation kinetics.
	
\begin{figure}
		\centering
		\includegraphics[width=0.45\textwidth]{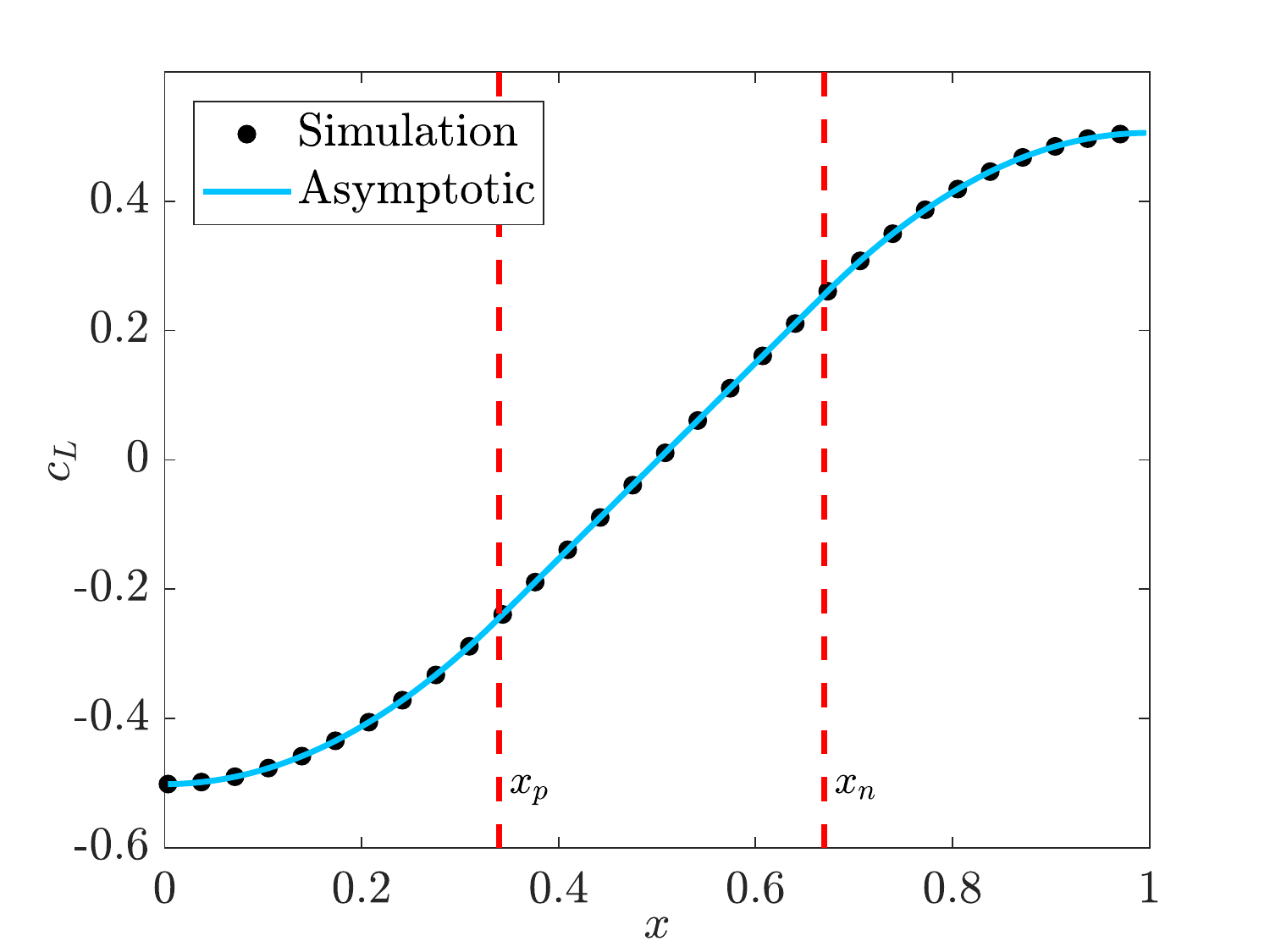}
		\caption{Analytical steady state profile \cref{eqn:clanal} for the electrolyte concentration compared to numerical simulation at $t=297$.}
		\label{fig:electrolyte}
	\end{figure}

	Numerical and asymptotic predictions of the steady-state concentration of lithium ions in the electrolyte are given in \cref{fig:electrolyte}. The simulation data is taken at the final time $t=297$.  However, the steady-state profile is numerically achieved within a few time steps consistent with the $\Ord{1}$ time analysis. The asymptotic prediction is given by \cref{eqn:clanal}.
	
\begin{figure}
		\centering
		\subfloat[Regional comparison.]
		{
			\label{subfig:cellass} \includegraphics[width=0.47\textwidth]{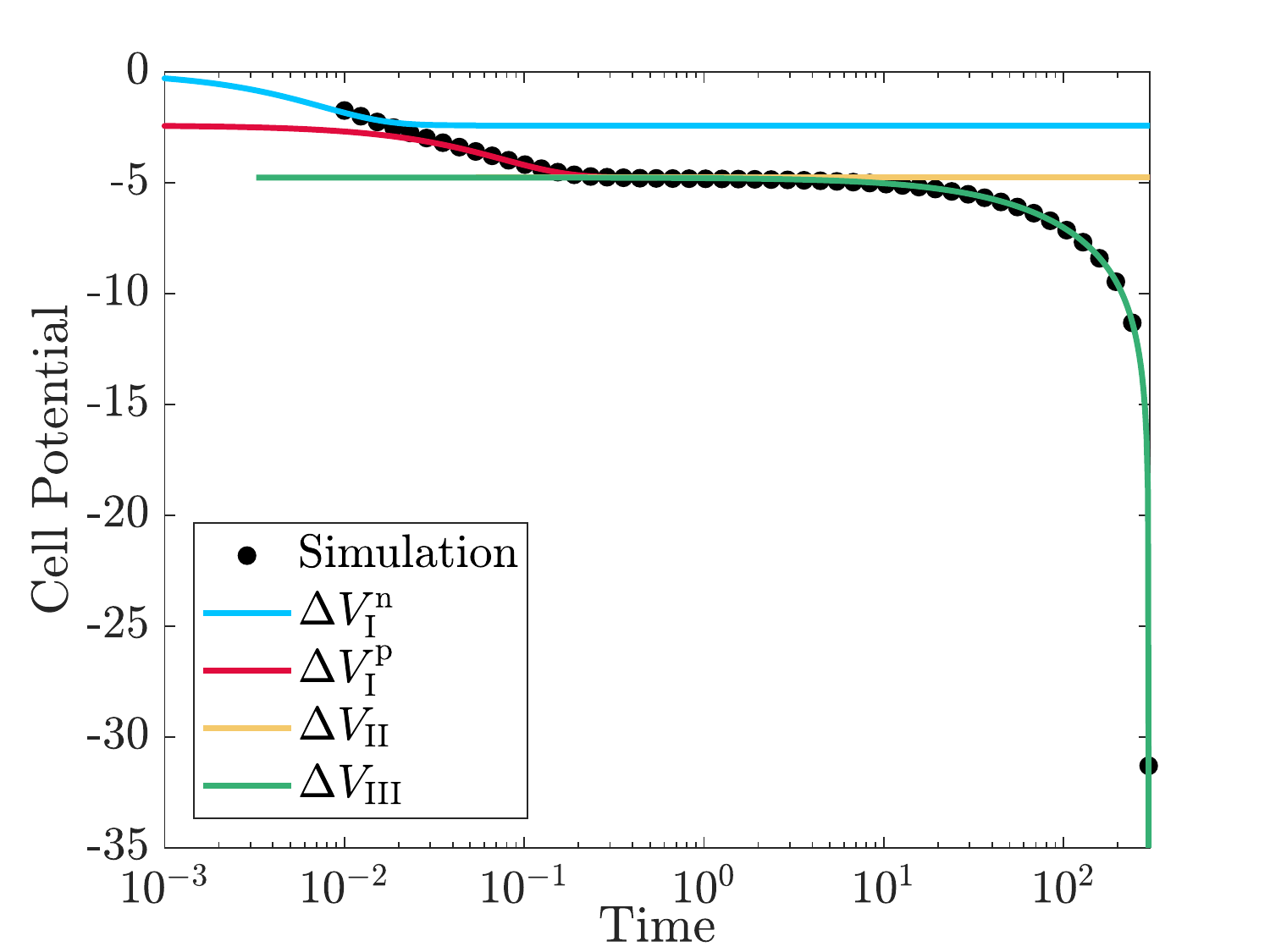}
		}
		\subfloat[Composite model comparison.]
		{
			\label{subfig:cellcomp} \includegraphics[width=0.47\textwidth]{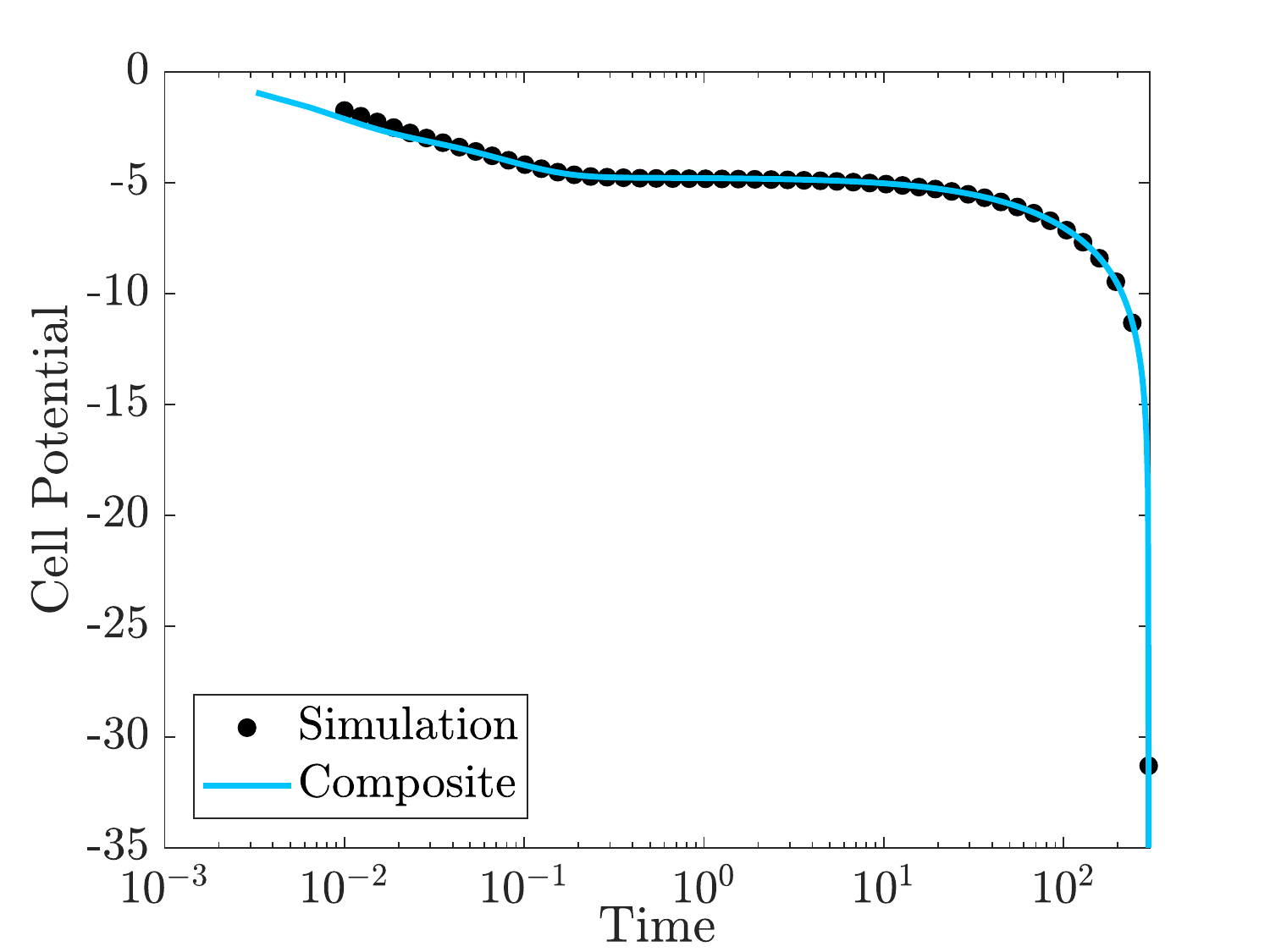}
		}
		\caption{Simulated discharge curve at 1C compared to the asymptotic cell potentials \cref{eqn:VI,eqn:VII,eqn:VIII,eqn:VIV}.  The time axis is presented on a logarithmic scale to emphasize the asymptotic regions in time where each of the mechanisms discussed in \cref{sec:ass1,sec:ass2,sec:ass3} dominate.  The composite solution in panel (b) is obtained by solving \cref{sys:compred}.} 
		\label{fig:cellvolt}
	\end{figure}

	We next plot the most relevant curve from an operational standpoint, the discharge curve.  This is a plot of the cell potential \cref{eqn:cellvolt} over the time span of discharge and captures the effects at all of the time regimes analysed.  The simulated data is compared to each of the asymptotic potentials \cref{eqn:VI,eqn:VII,eqn:VIII,eqn:VIV} in \cref{subfig:cellass}.  We also compare the simulated data to the composite reduced model \eqref{sys:compred} in \cref{subfig:cellcomp}.

\begin{figure}
		\centering
		\subfloat[$\mathcal{I}=10$.\label{subfig:I10}]
		{
			\includegraphics[width=0.45\textwidth]{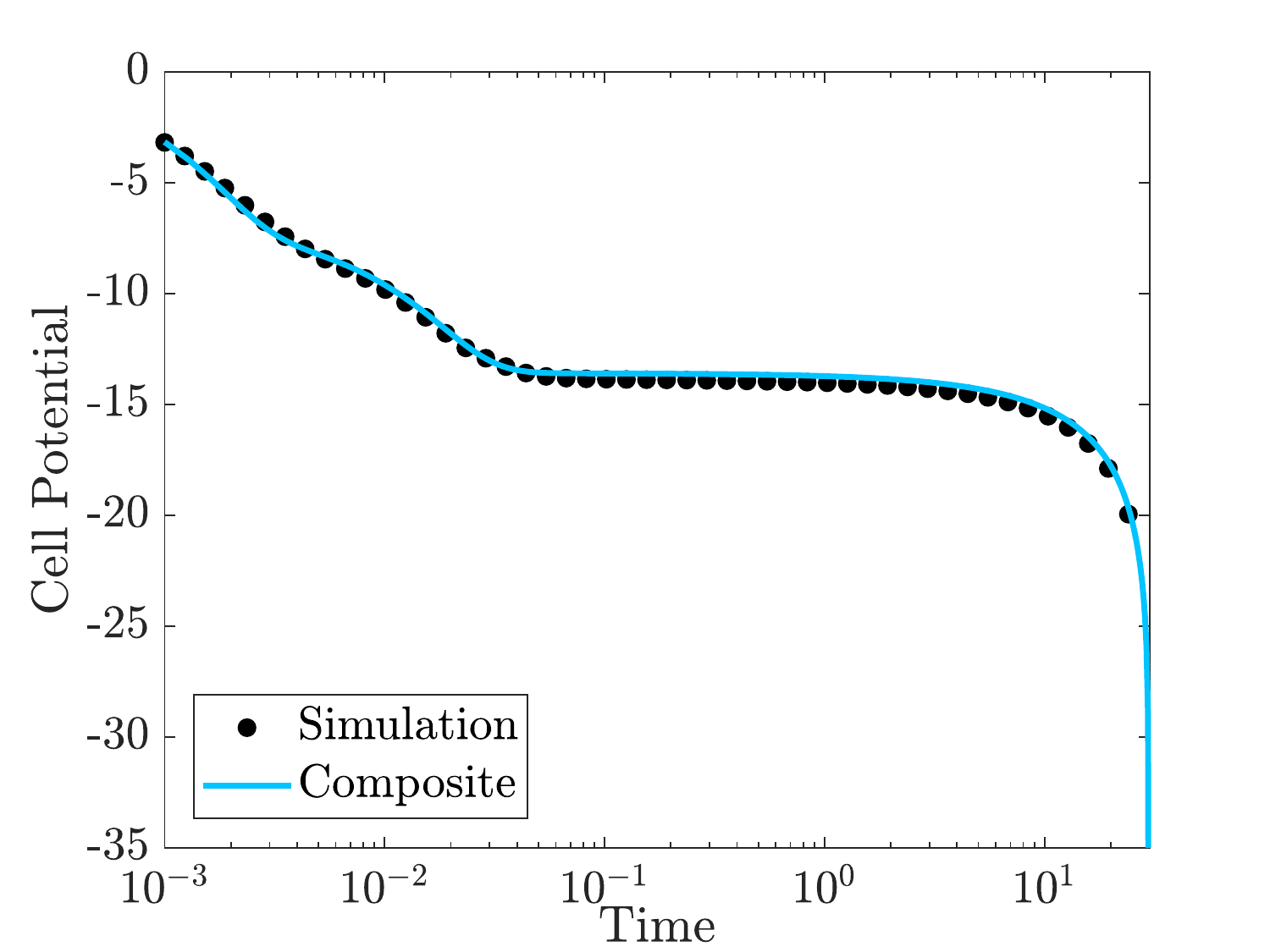}
		}
		\qquad
		\subfloat[$\mathcal{I}=100$.\label{subfig:I100}]
		{
			\includegraphics[width=0.45\textwidth]{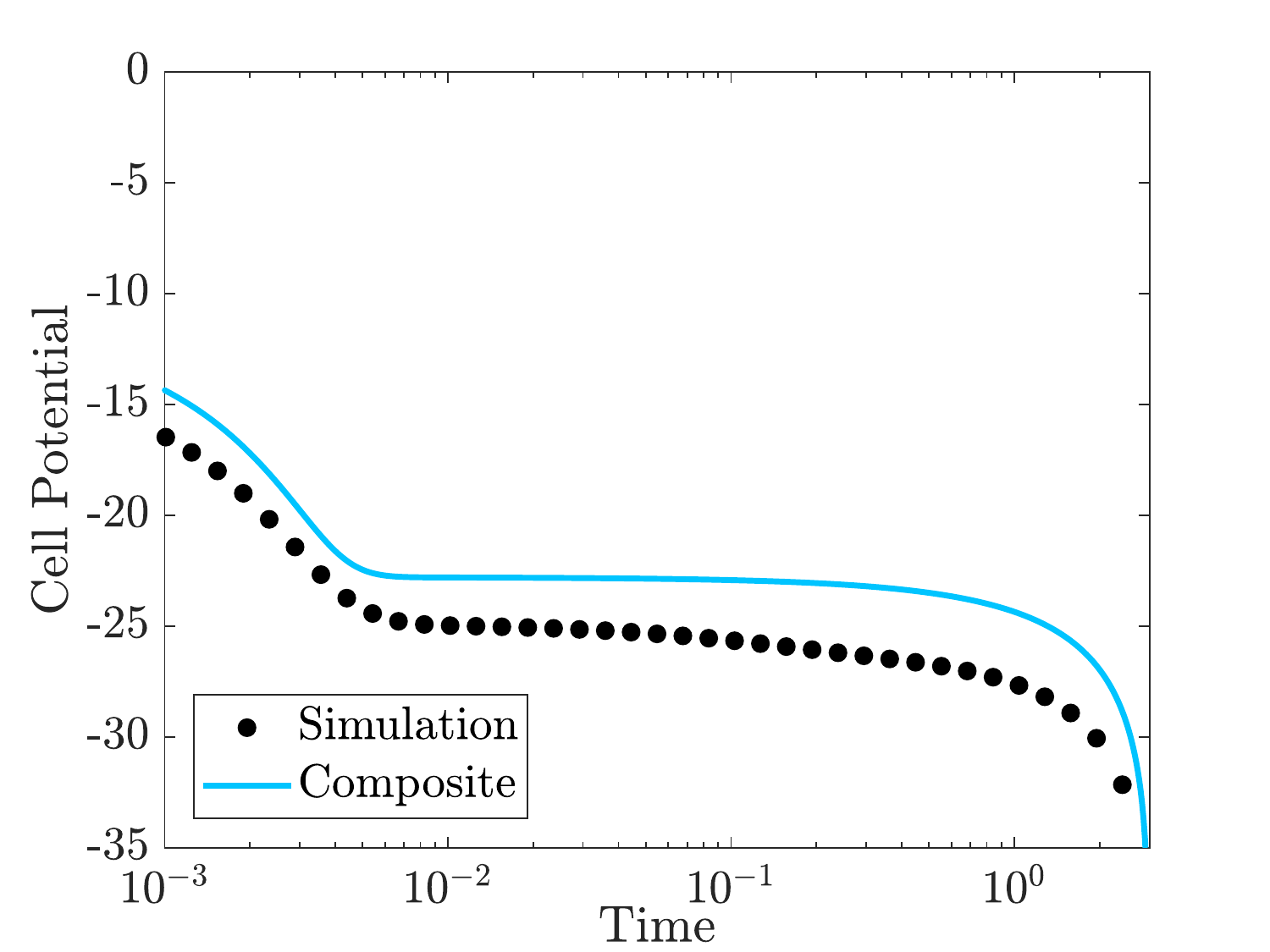}
		}
		\caption{Simulated discharge curves at 10C and 100C from numerical simulations of the full model (circles) and the composite reduced model (lines) given by \eqref{sys:compred}.}
		\label{fig:Icomp}
	\end{figure}

	For a given battery, the primary (dimensionless) parameter that can be varied is the C-rate, $\mathcal{I}$. Therefore, we demonstrate the robustness of the asymptotic reduction to this parameter in \cref{fig:Icomp} for high C-rates of 10 and 100, indicating fast discharge processes. We see that the quantitative agreement is excellent for the 10C case, but discrepancies occur at 100C.  However, the qualitative agreement that is observed at 100C indicates the persistence of the regimes identified by the asymptotic analysis, suggesting there has been no change in the dominant physical mechanisms taking place during battery discharge.
	
	The onset of discrepancies between the asymptotic and numerical solutions in \cref{fig:Icomp} can be understood by noticing that increasing $\mathcal{I}$ is equivalent to increasing the dimensional current $i_0$ which impacts several of the non-dimensional numbers. Thus, the failure of the asymptotic model around $\mathcal{I}=100$ is unsurprising because it leads to $\nui$ and $\gamma$ becoming $\Ord{1}$ in magnitude. 
	This has implications throughout the whole reduction which is based on (i) the concentration of electrolytic lithium $\cli$ not entering the leading-order Bulter--Volmer kinetics and (ii) the electric potentials being spatially uniform. Simulations for $\mathcal{I} \leq 10$ show excellent agreement with the reduced model.  Both high and low C-rate charges and discharges are important.  Low C-rate discharges allow for accurate measurements of open-circuit voltages \cite{Safari2011b}.  High C-rate charges are important for fast-charging mobile phones and electric vehicles.  However, it is known that these high rates can lead to battery degradation and capacity fade. Therefore, further modelling and analysis in this regime is warranted \cite{Choi2002}.
	
	\subsection{Comparison with experimental data}\label{sec:data}
	We demonstrate the effectiveness of our reduced model by comparing it to discharge data from a real battery obtained by Li \etal\cite{Li2014}. The dimensional parameters are provided in \cref{tab:params1} and \cref{tab:params2} of the supplementary material, and lead to $L/H = 2.0\tten{-3}$, $\Dn = 1.5\tten{-4}$, $\Dp=1.9\tten{-4}$, $\Da=1.8$, $\gamma=5.8\tten{-2}$, $\nusn=6.8\tten{-4}$, $\nusp=3.2\tten{-2}$, $\nue=2.1\tten{-2}$, $\Gn=4.9$, $\Gp=5.3$, $\Cn=7.4\tten{-5}$, $\Cp=5.1\tten{-3}$, $\xin=0.86$, $\xip=0.02$, $\deltap=2.39$, and $\deltan=4.5\tten{-2}$,
	which are of the presumed size for the asymptotic reduction in \cref{sec:asymptotics}. Therefore, the composite reduced model should sufficiently describe the battery being discharged. However, the theoretical open-circuit voltage \cref{eqn:butleretadef} is not used by Li \emph{et al}., who instead choose (dimensional) empirical formulae, $\mn{U}{\rm ref}{i}$, as functions of state of charge for a commercial battery with electrodes made of \ce{LiFePO_4} and graphite. The open-circuit voltage curves are shown in \cref{fig:OCV} with the corresponding empirical formula in \cref{sec:params} of the supplementary material. From our scaling, the state of charge, $\lambda$, can be related to the active solid lithium concentration via $\mnn{\lambda}{i}=\xii(1+\gamma\deltai\csi)$ and $\Ui$ is determined by $(R T_a/F) \log \Ui = \mn{U}{\rm ref}{i}(\xii)$.

	\begin{figure}
		\subfloat[Negative Electrode (graphite).\label{fig:OCVn}]
		{
			\includegraphics[width=0.45\textwidth]{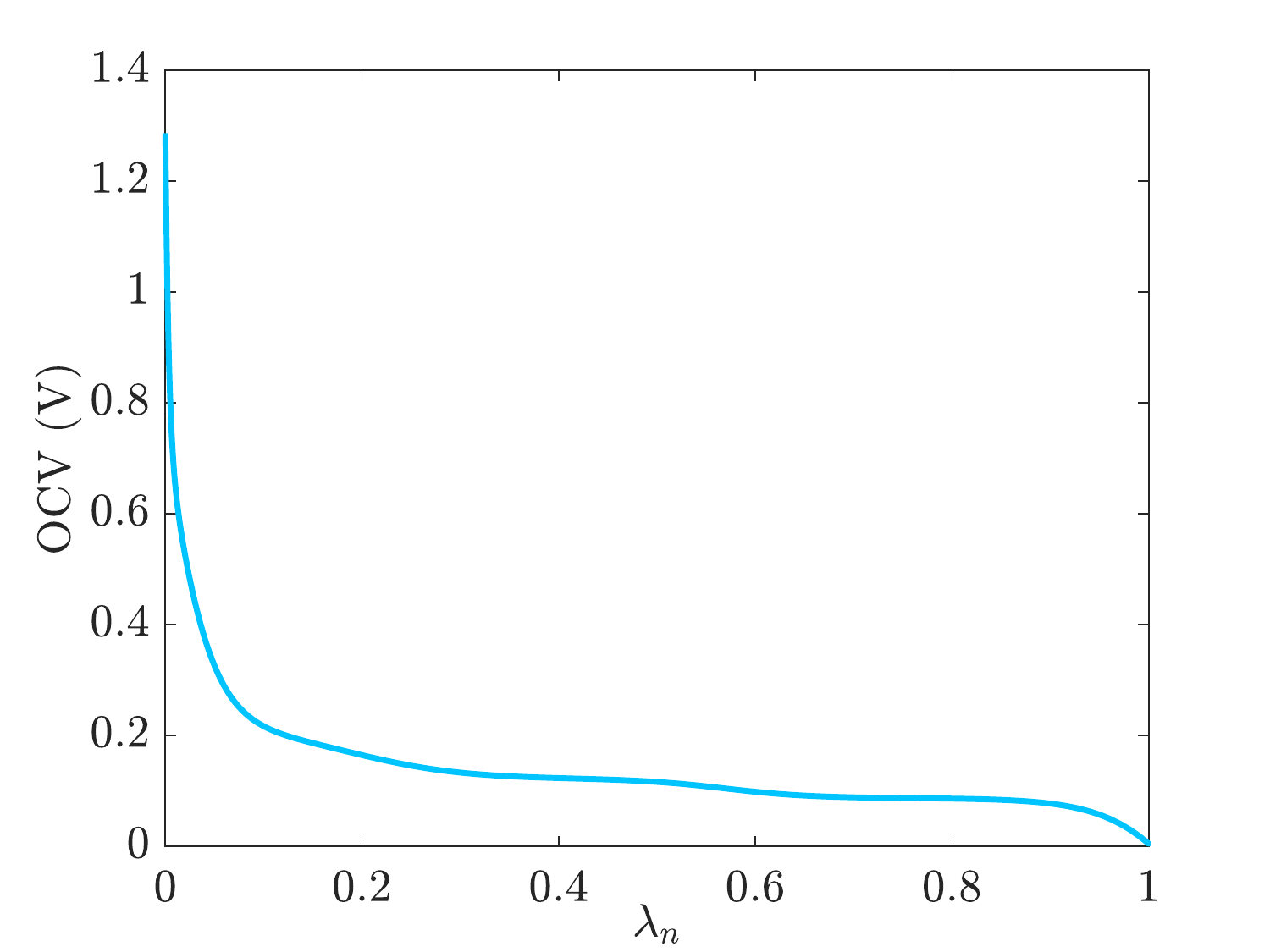}
		}
		\qquad
		\subfloat[Positive Electrode (LFP). \label{fig:OCVp}]
		{
			\includegraphics[width=0.45\textwidth]{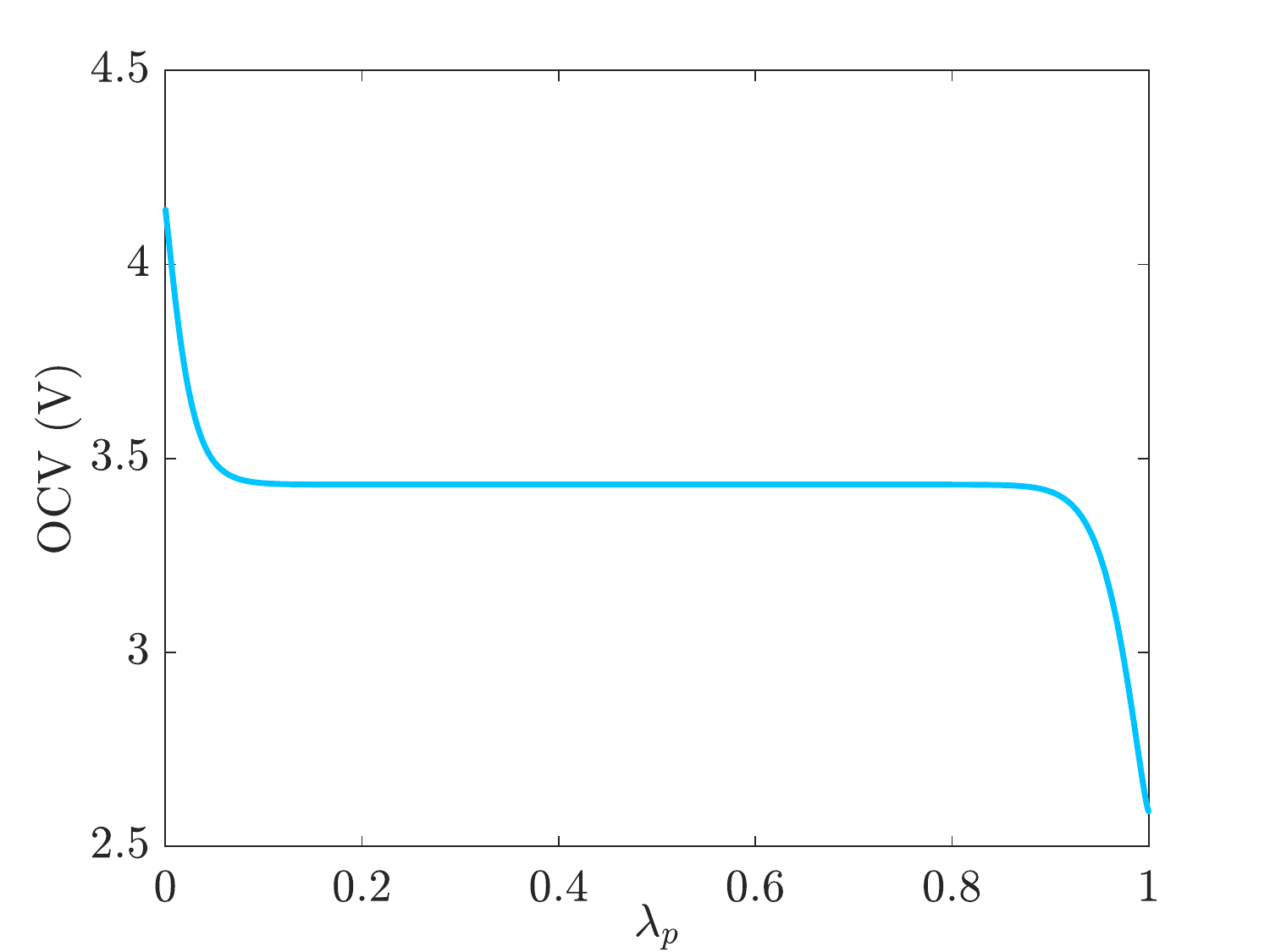}
		}
		\caption{Empirical open-circuit potentials used for a commercial lithium iron phosphate (LFP) battery from Li \etal\cite{Li2014}.}
		\label{fig:OCV}
	\end{figure}
	
\begin{figure}
		\centering
		\subfloat[Comparison between a numerical simulation of the composite reduced model \cref{sys:compred} (line), experimental battery discharge data at 2C (`X') from Li \etal\cite{Li2014}, as well as large-scale simulations of their P2D model (`$\cdot$').\label{fig:datacomp}]
		{
			\includegraphics[width=0.45\textwidth]{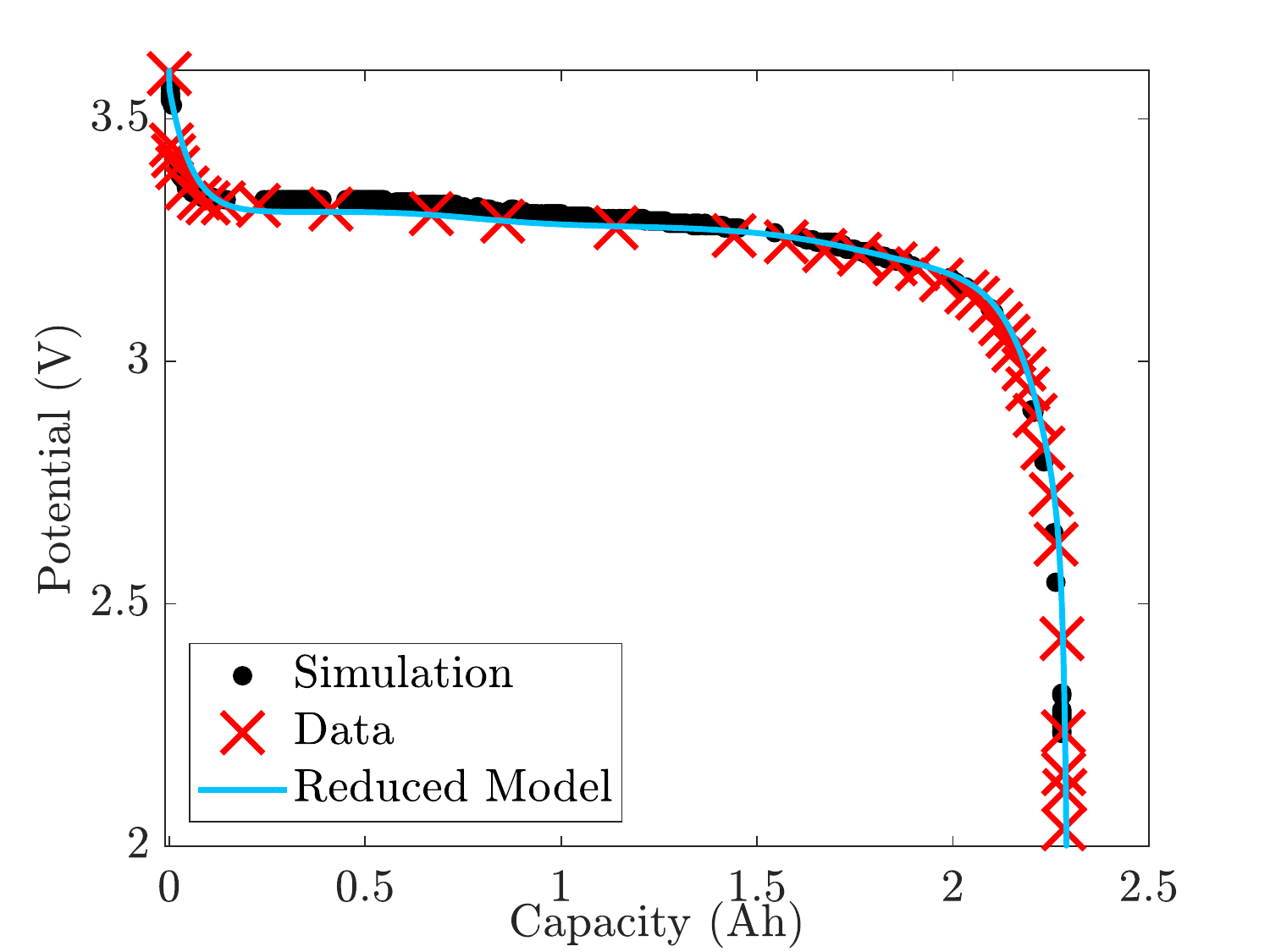}
		}
		\qquad
		\subfloat[Comparison between a numerical simulation of the composite reduced model \cref{sys:compred} at 0.1C (blue line), 1C (red line), and 3C (green line) with experimental battery discharge data (`X') from Safari and Delacourt\cite{Safari2011}.\label{fig:datacomp2}]
		{
			\includegraphics[width=0.45\textwidth]{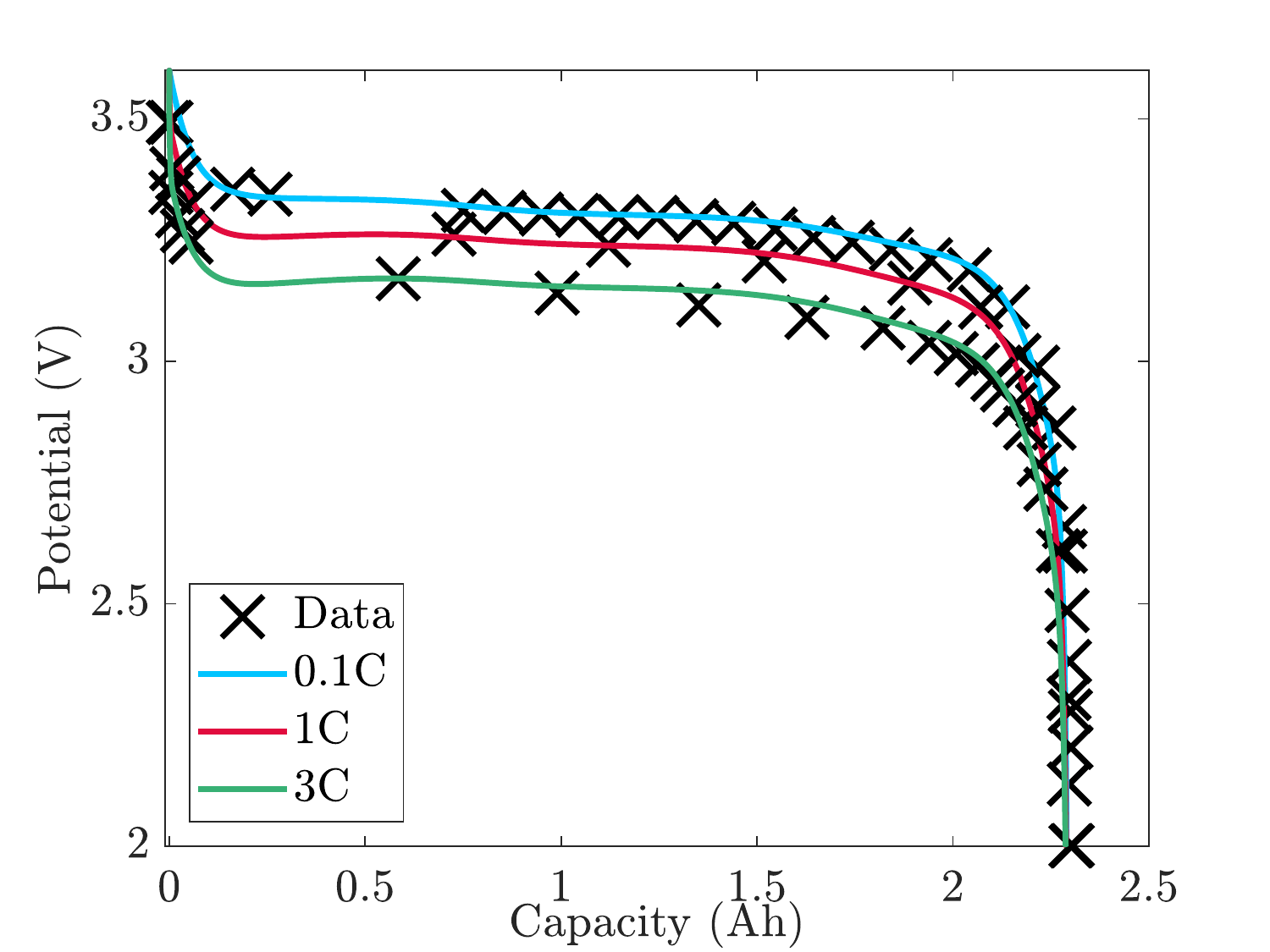}
		}
		\caption{}
		\label{fig:data_global}
	\end{figure}

	The composite reduced model \cref{sys:compred} is valid for all of the regimes and we solve it numerically using the implicit solver {\tt ode15s} in MATLAB until the state of charge of one of the electrodes decreases past zero or increases past one. This is equivalent to numerically finding the saturation/depletion times discussed in \cref{sec:ass3}. We compare our results to the data in Li \etal\cite{Li2014} for a 2C ($\mathcal{I}=2$) discharge rate in \cref{fig:datacomp} where excellent agreement is observed. An identical battery is used by Safari and Delacourt \cite{Safari2011} who provide discharge data at various C-rates. We compare \cref{sys:compred} to this data using the same parameters as before except with $\Cn=2.69\tten{-4}$, $\Cp=5.4\tten{-2}$, $\Gn=1.33$, and $\Gp=0.5$. These changes are due to Safari and Delacourt using different reaction rate constants for the kinetics compared to Li \etal The comparison between the reduced model and the data of Safari and Delacourt for discharge rates of 0.1C, 1C, and 3C is presented in \cref{fig:datacomp2}.  There is favourable agreement across all C-rates, with the accuracy of the model improving as the C-rate decreases. This demonstrates the feasibility of using the reduced model to predict discharge curves across a range of intermediate charging and discharging rates. The simulation results in each figure are rescaled dimensionally for appropriate comparison. Capacity is defined as the amount of charge used by the battery (in Ampere hours) and is scaled so that the time to full discharge corresponds to utilising the entire charge of the battery. This is done to normalise the discharge process as different C-rates correspond to different discharge times.
	
	It is important to point out that comparisons with experimental galvanostatic discharge curves only provides a means of assessing the validity of the reduced model in the third time regime. A more rigorous validation of the model would aim to probe the capacitance and diffusive regimes using experimental data from electrochemical impedance spectroscopy or sequential potentiostatic steps.
	
	\section{Discussion}\label{sec:results}
	The results in \cref{fig:csass,fig:electrolyte,fig:cellvolt,fig:Icomp} clearly demonstrate excellent agreement between the asymptotic theory and simulation.  The asymptotic reduction is simple and elegant due mostly to the spatial independence of the electric potential owing to $\nui\ll1$, which allows the problem to be decoupled.  We are able to show that the solid-phase lithium concentration is always a linear function of time, reproducing the numerical results of Li \etal\cite{Li2014}. Using parameters from this paper we were also able to demonstrate strong agreement between our reduced model and actual battery discharge data in \cref{fig:datacomp}. To compare to data, we numerically solved a pair of ordinary differential equations which is in contrast to the model simulated by Li \etal\cite{Li2014} (also plotted in \cref{fig:datacomp}) which uses twelve highly nonlinear partial differential equations. While algorithmic efficiency and optimisation can lead to fast solutions for a larger scale model of this type, a reduced model involving two simple differential equations requires much less sophistication to solve. Interestingly, the model solved by Li \etal uses the P2D approach discussed in \cref{sec:model}. Our reduced model also agrees favourably with their simulation results, validating that the volume-average approach used here is sufficient for matching discharge data under the presumed parameter size estimates. We demonstrate robustness of matching to discharge dynamics by also successfully comparing to data from Safari and Delacourt \cite{Safari2011} at 0.1C, 1C, and 3C discharge rates in \cref{fig:datacomp2}.
	
	As a counter example, our model is unable to predict the discharge curves measured by Srinivasan and Newman \cite{Srinivasan2004}, which shows that the capacity of their LFP electrode strongly depends on the discharge rate. There is a nearly 90\% reduction in capacity at a 5C discharge. Srinivasan and Newman compare their data to predictions from a ``shrinking core'' model, which accounts for phase separation by dividing each solid particle into an Li-rich shell and an Li-depleted core. They demonstrate that at least two particle sizes are required to fit the data. This indicates that the discharge characteristics depend on the microscale dynamics, a feature that is unlikely to be captured by the volume-average approach used here.
	
	An issue arises when comparing models to integrated data, such as the cell potential in \cref{fig:data_global} , which is that there is no spatial information to confirm or refute model predictions. For example, our volume-average approach predicts spatially homogeneous solid-phase concentration profiles in contrast to the P2D approach of Li \etal \cite{Li2014}, yet both models produce results that agree with experimental data. Interestingly, the P2D model of Ranom \cite[Sec.~2.5.5]{Ranom2014} in the limit of fast diffusion in the electrode particles also predicts that particles (de)lithiate at the same rate.
	Srinivasan and Newman \cite{Srinivasan2004} could accurately predict their discharge curves using a shrinking core model. However, phase-field simulations and experiments have since shown that shrinking core models do not provide realistic descriptions of (de)lithiation \cite{Cogswell2018,Lim2016} and obtaining reasonable agreement requires parameter values that contradict those obtained through experimental measurement.
	
	These differences in microscale modelling may not always be apparent when solely considering discharge dynamics, but may become more important for applications that aim to better understand the link between phase separation and battery behaviour, stress development in particles, active material utilisation, and degradation mechanisms.  Recent experiments involving X-ray microscopy and nuclear magnetic resonance imaging of \emph{in situ} lithium concentrations in solid and electrolyte \cite{Li2014Nature, Sethurajan2015,Krachkovskiy2018} may lead to more robust model predictions.  Nevertheless, the results presented here demonstrate that a reduced model derived from volume averaging can be an effective tool for accurately predicting battery operation, and one that offers substantial computational advantages over the P2D models commonly used in large-scale simulations \cite{Amiribavandpour2015, An2018, Li2014, Safari2011}.
	
	Contrasting desalination models \cite{Biesheuvel2010,Biesheuvel2011,He2018,Mirzadeh2014,Singh2018}, capacitive dynamics are seldom considered in LIB models. This is in spite of the fact that practical battery use may involve current pulses of short duration, where the battery response is dominated by capacitance effects \cite{Li2014, Ong1999}. Incorporating capacitive dynamics into large-scale numerical solvers must be done with care, as sophisticated time-stepping schemes are required to correctly capture rapid changes that occur on the capacitive time scale along with the normal operational changes that occur on larger time scales such as those associated with diffusion. 
	
	We have demonstrated that a volume-averaged model and its asymptotically reduced form are able to accurately predict battery behaviour. Since the asymptotic solutions are determined from a system of two ordinary differential equations, they can vastly speed-up prototyping as results can be quickly computed for a variety of parameters and compared to measured quantities. Battery designs that fail to fit with the model may indicate the importance of modelling physics which are not presented here. Indeed, as battery material research advances, the electronic and mechanical properties of electrodes will need to be integrated into electrochemical models.  For example, recent research \cite{Kennedy2017, Stokes2017} has shown that the structure of nanowire-based electrodes can have an important impact on battery capacity and electrolyte interactions. As new physics are introduced into models, the computational times of large-scale simulations will rapidly increase.  The use of simpler models obtained through a systematic reduction can make accurate computations more feasible, thereby accelerating the development of future battery technologies.
	
	\section{Conclusions}
	\label{sec:conclusions}
	Overall, we have considered a simple electrochemical model for lithium-ion batteries. Using the fact that reaction kinetics dominate electrical effects ($\nu_i\ll1$), we have shown that cell voltage behaviour can be understood through a sequence of asymptotic regimes which elucidate simple underlying physical processes. These asymptotic regimes are likely to persist should features such as concentrated solution theory, concentration-dependent parameters, and separate liquid and solid geometries be incorporated into the model. The simplicity of the asymptotically reduced model will make it an appealing tool for battery scientists and engineers. Despite the emphasis on battery application, we have maintained generality so that a similar problem reduction may be amendable to other electrochemical systems with comparable features.

	\section*{Acknowledgments}
	IRM and BRW would like to thank the Centre de Recerca Matem\`{a}tica for hosting them during the completion of this manuscript.
%%%%%%%%%%%%%%%%%%%%%%%%%%%%%%%%%%%%%%%%%
%%%%%%%%%%%%%%%%%%%%%%%%%%%%%%%%%%%%%%%%%
%%%%%%%%%%%%%%%%%%%%%%%%%%%%%%%%%%%%%%%%%	
\begin{appendix}
\section{Model Derivation}\label{sec:fullmod}
Traditionally, equations are presented in a volume-averaged form without derivation from the underlying microscopic equations with two exceptions being the work of Wang and Gu \cite{Wang1998} and Richardson \etal\cite{Richardson2012}.  For posterity then, we now present the full conservation of mass and charge model for each phase.

The volume averaging proceeds by first defining representative elementary volumes $\Omega=\Omega_\text{a} + \Omega_\text{ia}+\Omega_\text{e}$ containing domains of active solid material, inactive solid material, and the electrolyte respectively.  The microscopic model is then formulated in terms of equations which hold on each subdomain of the electrodes and separator.  Details on volume averaging including the conditions on selecting an appropriate representative volume can be found in Refs.~\cite{Whitaker1969, Bear1972, Fowler1997, Kaviany2012}.

\subsection{Electrode Model}
\noindent Lithium exists in a solid matrix of active material as intercalated particles which fit into the lattice spacing of the solid electrode material.  They diffuse through the active material until they reach the solid-liquid interface where current will cause an electron to leave and be carried by the electrode and a lithium ion will emerge into the liquid volume.  If an opposite current is applied then the process is reversed and lithium ions enter the solid as intercalated lithium and diffuse throughout.  The conservation of mass of a concentration of intercalated lithium, $\csi$ (mol m\unit{-3}), in the active material takes the form
\begin{align}
\pderiv{\csi}{t}=-\nabla\cdot\vecNsi;\qquad \vecNsi=-\mn{D}{a}{i}\nabla \csi,\label{eqn:csmass}
\end{align}
where $\mn{D}{a}{i}$ (m\unit{2} s\unit{-1}) is the diffusion coefficient of intercalated lithium.  The current in the active phase is given by Ohm's law,
\begin{align}
\vecisi=-\mn{\sigma}{a}{i}\nabla\Phisi,\label{eqn:ohm}
\end{align}
where $\vecisi$ (A m\unit{-2}) is the active phase current density, $\mn{\sigma}{a}{i}$ (S m\unit{-1}) is the electrical conductivity of the medium, and $\Phisi$ is the active phase potential (V).  Finally in the active phase, we impose conservation of charge, which leads to
\begin{align}
\nabla\cdot\vecisi=0.\label{eqn:solidneutral}
\end{align}
The equations for the active phase hold on each of the electrodes and parameters such as $\mn{D}{a}{i}$ and $\mn{\sigma}{a}{i}$ can be, and usually are, different for each of the two electrodes.

The liquid phase has two mobile charged species: the lithium ions, with concentration $\cli$ (mol m\unit{-3}), that are liberated from the solid and the anion, with concentration $\mn{c}{A}{i}$ (mol m\unit{-3}), that dissociate from the salt.  Assuming that the electrolyte fluid velocity is zero, conservation of mass of each species gives
\begin{align}
\pderiv{\mn{c}{j}{i}}{t}=-\nabla\cdot\mn{\vec{N}}{j}{i};\qquad \mn{\vec{N}}{j}{i}=-(\mn{D}{j}{i}\nabla \mn{c}{j}{i}+z_j\mu_jF\mn{c}{j}{i}\nabla\Phiei),\label{eqn:cimass}
\end{align}
where $z_j$ is the charge of the species, $\mu_j$ is the mobility (mol m\unit{2} J\unit{-1} s\unit{-1}), $F$ is Faraday's constant ($F\approx 96487$ C mol\unit{-1}), and $\Phiei$ is the electrolyte potential.  The electrolyte current is given by
\begin{align}
\veciei=F(\vecNli-\mn{\vec{N}}{A}{i}),\label{eqn:ie}
\end{align}
and charge neutrality in the electrolyte states that
\begin{align}
\sum_iz_i\mn{c}{i}{i}=0.
\end{align}
This results in
\begin{align}
\nabla\cdot\veciei=\nabla\cdot(\vecNli-\mn{\vec{N}}{A}{i})=0.\label{eqn:liquidneutral}
\end{align}
A secondary consequence of charge neutrality is that $\cli=\mn{c}{A}{i}$ which, due to a global conservation of mass, must also equal the concentration of the solvent.  In writing the flux $\mn{\vec{N}}{i}{i}$ in the form \cref{eqn:cimass}, we have implicitly assumed ideal conditions such as an infinitely-dilute electrolyte.  Otherwise, components such as the electrolyte potential are difficult to define and instead one considers a multi-component mass transfer such as in \cite[Chapter 12]{Newman2004}.  Furthermore, the dilute assumption is convenient for selecting the correct scales.

We will now consider boundary conditions between the active and liquid phase. Anions cannot enter the active solid,
\begin{align}
\mn{\vec{N}}{A}{i}\cdot\vec{n}_s=0,\qquad \vec{x}\in\partial\Omega_s.
\end{align}
Secondly, intercalated lithium that leaves the active solid phase enters the electrolyte phase and so there is a global conservation of mass and therefore, at the boundary, the mass fluxes must satisfy,
\begin{align}
\vecNli\cdot\vec{n}_s=\vecNsi\cdot\vec{n}_s,\qquad \vec{x}\in\partial\Omega_s.\label{eqn:cflux}
\end{align}
However, we still need to provide a condition for the mass flux out of the solid itself and we do this by analysing the surface charge.  While electroneutrality occurs in the bulk of each phase, there are two contributing sources to boundary charge transfer.  Firstly there are the electrochemical reactions at the electrode surface which transforms the intercalated lithium to ions and secondly there is an electric double layer which forms near the electrode surface that induces current because of a change in surface charge.  We are considering electrode reactions of the form
\begin{align}
\ce{ILi
	<=>[charge][discharge]
	\ce{I} + \ce{Li^+} + \ce{e^-},
}
\end{align}
where I is the intercallating material holding the lithium.  In this case, Faraday's law dictates that the chemical reactions at the surface are \cite[page 374]{Newman2004}: 
\begin{align}
F\mn{\vec{N}}{j}{i}^f\cdot\vec{n}_s=\mnn{g}{i},\qquad \vec{x}\in\partial\Omega_s,
\end{align}
where $\mn{\vec{N}}{j}{i}^f$, with $j = a, L$, are the Faradaic mass fluxes and $\mnn{g}{i}$ is the electrochemical current generated by the reactions.  Following, for example, Newman and Tiedemann \cite{Newman1975} and Newman and Thomas-Alyea \cite[pg. 522]{Newman2004}, the surface charge conservation satisfies
\begin{align}
\pderiv{\mn{q}{a}{i}}{t}=F\vecNsi^f\cdot\vec{n}_s-F\vecNsi\cdot\vec{n}_s,\qquad \vec{x}\in\partial\Omega_s,
\end{align}
where $\mn{q}{a}{i}$ is the solid surface charge density. Due to electroneutrality, $\mn{q}{a}{i}=-F\mn{c}{\Gamma}{i}$, where $\mn{c}{\Gamma}{i}$ is the liquid surface charge density of lithium ions.  If we define $\mn{C}{\Gamma}{i}$ (F m\unit{-2}) as the capacitance per unit area then the solid surface charge satisfies $\mn{q}{a}{i}=-\mn{C}{\Gamma}{i}(\Phisi-\Phiei)$ and the mass flux becomes
\begin{align}
\begin{aligned}
F\vecNsi\cdot\vec{n}_s=&F\vecNsi^f\cdot\vec{n}_s+\mn{C}{\Gamma}{i}\pderiv{}{t}(\Phisi-\Phiei),\\
F\vecNli\cdot\vec{n}_s=&F\vecNli^f\cdot\vec{n}_s+\mn{C}{\Gamma}{i}\pderiv{}{t}(\Phisi-\Phiei),
\end{aligned}
\qquad \vec{x}\in\partial\Omega_s,
\end{align}
where we note that the flux condition \cref{eqn:cflux} is satisfied.  The closure condition is that the mass flux out of the solid must provide the solid  current density,
\begin{align}
(F\vecNsi-\vecisi)\cdot\vec{n}_s=0,\label{eqn:ibc}
\end{align}
and this must also be equal to the electrolyte current in order to have continuity of current densities,
\begin{align}
(\veciei-\vecisi)\cdot\vec{n}_s=0.
\end{align}

\subsection{Separator Model}
The separator, like the electrode, is also a porous media but with the caveat that there is no mass transport in the solid phase which exists to electrically insulate the electrodes from one another.  Therefore, we can write down conservation of mass and electroneutrality as
\begin{subequations}
	\begin{align}
	\pderiv{\mn{c}{i}{s}}{t}=&-\nabla\cdot\vec{N}_i, \\
	\mn{\vec{N}}{i}{s}=&-(D_i\nabla \mn{c}{i}{s}+z_i\mu_iF \mn{c}{i}{s}\nabla\Phies),\\
	\vecies=&F(\vecNls-\mn{\vec{N}}{A}{s}),\\
	\nabla\cdot\vecies=&0.
	\end{align}
\end{subequations}
The boundary conditions for mass flux are,
\begin{align}
\vecNls\cdot\vec{n}_{s}=\mn{\vec{N}}{A}{s}\cdot\vec{n}_{s}=0, \qquad \vec{x}\in\partial\Omega_{s}.
\end{align}

\subsection{Boundary Conditions}\label{sec:BC}
We now need to apply boundary conditions to the full model geometry in \cref{fig:setup}.  Firstly, at the separator-electrolyte boundaries, we will enforce continuity of concentration and fluxes in the liquid phase,
\begin{align}
\begin{aligned}
\mn{c}{i}{i}=&\mn{c}{i}{s},\\
(\mn{\vec{N}}{i}{i}-\mn{\vec{N}}{i}{s})\cdot\vec{n}=&0,
\end{aligned}
\qquad x=x_p\textrm{ and }x=x_n,
\end{align}
and for the active solid phase that there is no flux of intercalated lithium into the separator,
\begin{align}
\begin{aligned}
\vecNsi\cdot\vec{n}=&0,
\end{aligned}
\qquad x=x_p\textrm{ and }x=x_n.
\end{align}
The normal vector here refers to the outer normal of the macroscale area in \cref{fig:setup}.  We also stipulate that
\begin{align}
\vecisi\cdot\vec{n}=0,\quad(\veciei-\vecies)\cdot\vec{n}=0,\qquad x=x_p\textrm{ and }x=x_n,
\end{align}
so that solid carries no current as it leaves the electrodes and that the electrolyte current densities are continuous across the interface.  At $x=0$ we will apply a current to the solid phase only,
\begin{align}
\vecisp\cdot\vec{n}=i_{\rm app},\quad \veciep\cdot\vec{n}=0;\qquad x=0,\label{SM-bc:iapp}
\end{align}
while at the edge of the other electrode we apply a grounding condition and also stipulate that the current is carried entirely by the solid,
\begin{align}
\Phien=0,\quad\vecien\cdot\vec{n}=0;\qquad x=L.
\end{align}
The applied current $i_{\rm app}$ appearing in \cref{bc:iapp} is often given in terms of the C-rate which is a measure of how much a battery's capacity has been used in one hour.  For example, if a battery is rated as 1 Ah, the standard unit of capacity, then a 1C rate would correspond to a current of 1 A while a 0.5C and 2C rate would correspond to 0.5 A and 2 A and a charge/discharge time of 2 hours and 30 minutes respectively.  Taking this into consideration we will define the applied current as,
\begin{align}
i_{\rm app}=\mathcal{I}i_0\label{eqn:Cratedef}
\end{align}
with $\mathcal{I}$ the C-rate and $i_0$ (A m\unit{-2}) the normal operating current density provided by the device.  This current is given by
\begin{align}
i_0=\frac{I_{\rm app}}{A_{\rm cell}},\label{eqn:i0def}
\end{align}
where $I_{\rm app}$ is the draw current at a discharge rate of 1C and $A_{\rm cell}$ is the area of the electrode.

We define the volumes of each $P$, $S$, and $N$ in \cref{fig:setup} as $\Omega_P$, $\Omega_S$, and $\Omega_N$ respectively and then let the global external boundary be denoted $\partial\Omega_{P\cup S\cup N}$.  On this boundary we will apply no mass flux of any species,
\begin{align}
\mn{\vec{N}}{i}{i}\cdot\vec{n}=0;\qquad \vec{x}\in\partial\Omega_{P\cup S\cup N}.
\end{align}

\subsection{Volume averaging}

The model as posed can now be volume averaged.  We define the volume average and intrinsic volume average of a quantity $\psi$ as
\begin{align}
\avg{\psi}=\frac{1}{V}\int_\Omega\psi\diff V,\qquad\avg{\psi}^i=\frac{1}{V_i}\int_{\Omega_i}\psi\diff V
\end{align}
respectively, where $V_i$ is the volume of domain $\Omega_i$.  The equations that result are 
\begin{subequations}\label{sys:avgmodel}
	\begin{align}
	\pderiv{}{t}(\phisi \csi)&=\nabla\cdot(\phisi \mn{D}{a}{i}\nabla \csi)+\frac{1}{F}\nabla\cdot(\phisi \vecisi), \label{eqn:csiavg}\\
	\vecisi&=-\mn{\sigma}{a}{i}\nabla\Phisi,\\
	\nabla\cdot(\phisi\vecisi)&=-\mnn{a}{i}\left(\gibar+\mn{C}{\Gamma}{i}\pderiv{}{t}(\Phisi-\Phiei)\right),\label{eqn:divisavg}
	\end{align}
	for the (active) solid-phase of the electrodes,
	\begin{align}
	\pderiv{}{t}(\phiei \cli)&=\nabla\cdot(\phiei D_L\nabla \cli +\phiei\mu_LF\cli\nabla\Phiei)+\frac{1}{F}\nabla\cdot(\phiei\veciei),\label{eqn:avgci}\\
	\veciei&=F(\vecNli-\mn{\vec{N}}{A}{i}),\\
	\nabla\cdot(\phiei\veciei)&=\mnn{a}{i}\left(\gibar+\mn{C}{\Gamma}{i}\pderiv{}{t}(\Phisi-\Phiei)\right),\label{eqn:divieavg}
	\end{align}
	for the liquid-phase of the electrodes, and
	\begin{align}
	\pderiv{}{t}(\phies \cls )&=\nabla\cdot(\phies D_L \nabla\cls +\phies \mu_LF\cls\nabla\Phies),\\
	\vecies &=F(\vecNls-\mn{\vec{N}}{A}{s}),\\
	\nabla\cdot(\phies \vecies)&=0,
	\end{align}
\end{subequations}
for the separator. The molar fluxes are given by
\begin{align}
\mn{\vec{N}}{j}{i}&=-D_j\nabla \mn{c}{j}{i}-z_j\mu_jF\mn{c}{j}{i}\nabla\Phiei. \label{eqn:vecN}
\end{align}
New parameters are introduced to the equations through volume averaging: $\phisi$ is the volume fraction of active solid material, $\phiei$ is the volume fraction of the electrolyte which also corresponds to the porosity of the electrode, defined as the ratio  of  the  electrolyte  volume  to  the  total  volume. Thus, $1-\phiei$ give the total volume fraction of the solid including both active and inactive materials. The parameter $\mnn{a}{i}$ is the specific area of active electrode material per unit volume, $\mnn{a}{i}=\mn{A}{ae}{i} / V$, where $A_{ae}$ is the surface area of the interface formed between active solid material and the electrolyte.

Adding \cref{eqn:divisavg} and \cref{eqn:divieavg} leads to
\begin{align}
\nabla\cdot(\phisi \vecisi+\phiei\veciei)=0,\label{SM-eqn:phaseneutral}
\end{align}
which is equivalent to (phase-averaged) conservation of charge.

In deriving these equations, we have assumed that the variation with respect to the volume average is zero so as to not pick up additional anisotropic tensor terms.  Secondly, each variable is intrinsically volume averaged over the active solid material which is in slight contrast to other literature where currents are left as volume averages over the entire volume.  These two averages can easily be connected by the relation,
\begin{align}
\avg{\vecisi}=\phisi \avg{\vecisi}^a.
\end{align}

The volume averaged boundary conditions are 
\begin{subequations}
	\begin{alignat}{2}
	\cli - \cls &=  0, &\quad  x&=x_p,\,x_n; \label{bc:cl_cont} \\
	\Phiei - \Phies &=  0, &\quad  x&=x_p,\,x_n;\\
	(\phiei \nabla \cli - \phies \nabla \cls)\cdot \vec{n} &=  0, &\quad  x&=x_p,\,x_n;\\
	(\phiei \nabla \Phiei - \phies \nabla \Phies)\cdot \vec{n} &=  0, &\quad  x&=x_p,\,x_n;\\
	\nabla \csi \cdot\vec{n}&=0, &\quad  x&=x_p,\,x_n; \\
	\vecisi\cdot\vec{n}&=0, &\quad  x&=x_p,\,x_n,\\
	\phisp\vecisp\cdot\vec{n}&=-\mathcal{I}i_0, &\quad x &= 0,\\
	\Phien&=0, &\quad x&=L,\\
	\nabla \cli \cdot\vec{n} &= 0, &\quad x &= 0, L,\\
	\nabla \csi \cdot\vec{n} &= 0, &\quad x &= 0, L,
	\end{alignat}
\end{subequations}
where some simplifications have been made. We have not included boundary conditions on the top and bottom of the cell as these will be unnecessary after an asymptotic reduction to one dimension.  The initial concentrations are assumed to be spatially uniform and given by $\csi(\vec{x},0)=\mn{c}{a0}{i}$ and $\cli(\vec{x},0)=c_{L0}$. The electric potentials satisfy $\Phiei(\vec{x},0)=0$ and $\Phisi(\vec{x},0)=(RT_a/F)\mnn{U}{i}$ with $\mnn{U}{i}$ defined as the non-dimensional open-circuit potential. The initial potential allocation is consistent with the grounding condition on $\Phien$.

\subsection{Reaction Kinetics}\label{sec:kinetics}
We have yet to prescribe the electrochemical kinetics that model the reaction currents $\gibar$ at the solid-electrolyte interfaces. Following Refs.~\cite{Doyle1993, Fuller1994, Newman2004, Newman1975}, we will use Butler--Volmer type kinetics, which have the form
\begin{align}
\gibar=\ji\left(\exp\left[\frac{(1-\betai)F}{RT_a}\etai\right]-\exp\left[\frac{-\betai F}{RT_a}\etai\right]\right),\label{eqn:fullBV}
\end{align}
where $R$ is the ideal gas constant, $T_a$ is the ambient temperature, and $\mnn{\eta}{i}$ is the surface overpotential \cite[page 211]{Newman2004},
\begin{align}
\etai=\Phisi-\Phiei-\mnn{U}{i}.\label{eqn:eta_def}
\end{align}
Each exponential term in \eqref{eqn:fullBV} represents a contribution of current both into and out of the electrode, referred to as anodic and cathodic reaction currents. The parameter $0 < \mnn{\beta}{i} < 1$ is a symmetry factor and represents the possibility that one reaction current direction is favoured over another. Furthermore, $j_{0,i}$ is the exchange current density and we will take it to have the form,
\begin{align}
\ji=F\mn{K}{a}{i}^{\betai}\mn{K}{L}{i}^{1-\betai}\csi^{\betai}\left(\frac{\csimax-\csi}{\csimax}\right)^{1-\betai}\cli^{1-\betai},\label{eqn:j0}
\end{align}
where $\mn{K}{j}{i}$ (m s\unit{-1}) are the heterogeneous reaction constants.  This has been adapted from Ref.~\cite[page 212]{Newman2004} whereby a carrying-capacity term has been introduced to represent the maximal density $\csimax$ of intercalated lithium allowed into the solid phase.  We do not include a term for the maximal electrolyte concentration under the notion that the capacity of the solid phase will be reached first and also because infinitely-dilute solution theory was assumed.  Using \eqref{eqn:j0} for the Butler--Volmer kinetics leads to the following definition of the open-circuit potential \cite[page 211]{Newman2004}:
\begin{align}
\mnn{U}{i}=\frac{RT_a}{F}\log\left[\frac{\mn{K}{L}{i}\cli(\csimax-\csi)}{\mn{K}{a}{i}\csi\csimax}\right].
\label{eqn:Ui}
\end{align}

By prescribing Butler--Volmer kinetics of the form \cref{eqn:fullBV}, we are assuming the that the electric double layer between the solid and electrolyte is in the Helmholtz limit. This limit assumes a single capacitive Stern layer of counter-ion charge adheres to the surface of the electrode. A more general model of the electric double layer considers the interface between the solid electrode matrix and the bulk of the electrolyte as having two contributions: a Stern layer and a stagnant diffusion layer. This second layer is a diffusive electrolyte regime experiencing a weaker electrostatic response to the surface charge. Overall, the difference in potential between the solid and bulk electrolyte is
\begin{align}
\mnn{\Phi}{s}-\mnn{\Phi}{e}=\Delta\mnn{\Phi}{D}+\Delta\mnn{\Phi}{S},
\end{align}
where $\Delta\mnn{\Phi}{D}$ is the potential drop across the stagnant diffusion layer and $\Delta\mnn{\Phi}{S}$ is the drop across the Stern layer. By assuming two regimes for the electric double layer, it is important to recognize that the Faradaic reactions between the solid and electrolyte are occurring across the Stern layer and not across the entire electrolyte. This leads to the Frumkin correction to the Butler--Volmer reaction kinetics ($\gibar^{\rm F}$) given by,
\begin{align}
\gibar^{\rm F}=F\mn{K}{a}{i}\csi\exp\left(\frac{(1-\betai)F}{RT_a}\Delta\mnn{\Phi}{S}\right)-F\mn{K}{L}{i}\cli^{\rm S}\left(\frac{\csimax-\csi}{\csimax}\right)\exp\left(-\frac{\betai F}{RT_a}\Delta\mnn{\Phi}{S}\right),
\end{align}
where $\cli^{\rm S}$ is the concentration of lithium in electrolyte reacting at the Stern layer.  The active solid lithium does not exist in the liquid phase and so its concentration is unaffected. We relate $\cli^{\rm S}$ to the bulk value through the Boltzmann equation, $\cli^{\rm S}=\cli\exp\left(-\frac{F}{RT_a}\Delta\mnn{\Phi}{D}\right)$ which along with simplifying \eqref{eqn:Ui} yields,
\begin{align}
\gibar^{\rm F}=F\mn{K}{a}{i}\csi\exp\left((1-\betai)\frac{F}{RT_a}\Delta\mnn{\Phi}{S}\right)\left(1-\exp\left(-\frac{F}{RT_a}\etai\right)\right),
\end{align}
where $\etai$ is still given by \cref{eqn:eta_def}.

An interpretation of the Helmholtz limit is that the Stern layer is much larger than the stagnant diffusion layer, effectively setting $\Delta\mnn{\Phi}{D}$ to zero. In this case $\Phisi-\Phiei=\Delta\mnn{\Phi}{S}=\etai+\mnn{U}{i}$ and using \cref{eqn:Ui} then $\gibar^{\rm F}$ reduces to $\gibar$ in \eqref{eqn:fullBV}, which we will consider for simplicity.

\subsection{Non-dimensionalisation}\label{sec:nondim}
The model is written in dimensionless form by introducing characteristic scales for all of the variables. These scales are obtained by considering the physics of battery operation. We define a common porosity $\phien=\phiep=\phie$ for the two electrodes for simplicity. 

The coordinates are written in terms of the cell length and height by letting $x = L x'$, $x_i=L x_i'$, and $y = H y'$, where primes are used to denote dimensionless quantities. Battery operation requires a continuous flow of lithium ions between the electrodes. The dominant mechanism of lithium transport in the separator is diffusion through the electrolyte (this will be verified below). Thus, time is non-dimensionalised using the time scale of lithium diffusion in the electrolyte, $t = (L^2 / D_L) t'$. The normal operating current $i_0$ defines a natural scale for the current densities in the model. Thus we write $\vecisi = i_0 \vecisi'$ and $\veciei = i_0 \veciei'$. The concentrations are written in terms of the deviation from their initial value using a characteristic scale $\Delta c = (i_0 L)/(F D_L)$ that captures the change in composition due to electrochemical reactions, leading to $\csi-\mn{c}{a0}{i} = (\Delta c) \csi'$ and $\cli-c_{L0} = (\Delta c) \cli'$. The applied current at the electrodes drives the electrochemistry which, in turn, sets the scale for the electric potential through the Butler--Volmer kinetics \cref{eqn:j0}. The electric and open-circuit potentials are therefore written as $\Phisi = (R T_a / F)[\log \Ui +  \Phisi']$, $\Phiei = (R T_a / F)\Phiei'$, $\mnn{U}{i} = (R T_a / F)[\log \Ui + \mnn{U}{i}']$ and $\etai = (R T_a / F) \etai'$, which make the exponents in the Butler--Volmer kinetics $\Ord{1}$ in magnitude. The combination $(R T_a / F) \log \Ui$ corresponds to the initial value of the dimensional open-circuit voltage and $\Ui$ will be defined below. The electrochemical current $\gibar$ is written as $\gibar = \mn{g}{0}{i} \gibar'$, where $\mn{g}{0}{i}$ is defined as
\begin{align}
\mn{g}{0}{i}&=F\mn{K}{a}{i}^{\betai}\mn{K}{L}{i}^{1-\betai}\mn{c}{a0}{i}^{\betai}\left(\frac{\csimax-\mn{c}{a0}{i}}{\csimax}\right)^{1-\betai}\mn{c}{L0}{i}^{1-\betai},\label{eqn:g0def}
\end{align}
which comes from non-dimensionalising the Butler--Volmer kinetics \cref{eqn:j0}. 

Assuming that the porosity of each domain is constant, the dimensionless bulk equations for the active solid components of the electrodes are given by (upon dropping the primes)
\subeq{
	\label{SM-sys:electrode_solid}
	\begin{align}
	\pderiv{\csi}{t}&=\Di\nabla^2\csi +\nabla\cdot\vecisi, \label{SM-eqn:nd_csi}\\
	\mn{\nu}{a}{i} \vecisi&=-\nabla\Phisi,\label{SM-eqn:ohma}\\
	\phisi\nabla\cdot \vecisi &=-\Gi \left(\gibar +\Ci \pderiv{}{t}(\Phisi-\Phiei)\right),
	\end{align}
}
where the gradient has been redefined as
\begin{align}
\nabla\equiv\left(\pderiv{}{x},\alpha\pderiv{}{y}\right)\label{eqn:newgrad}
\end{align}
and the dimensionless parameters
\begin{align}
\alpha = \frac{L}{H}, \quad
\Di = \frac{\mn{D}{a}{i}}{D_L}, \quad
\mn{\nu}{a}{i} = \frac{i_0 L F}{R T_a \mn{\sigma}{a}{i}}, \quad
\Gi = \frac{\mnn{a}{i}\mn{g}{0}{i}L}{i_0}, \quad
\Ci = \frac{\mn{C}{\Gamma}{i} R T_a D_L}{\mn{g}{0}{i} F L^2},
\label{eqn:nondim_1}
\end{align}
denote the aspect ratio of the cell ($\alpha$), the ratio of lithium diffusivity in the active solid to the diffusivity in the separator electrolyte ($\Di$), the relative electrical resitivity of the electrodes ($\mn{\nu}{a}{i}$), the dimensionless scale of the electrochemical current ($\Gi$), and the dimensionless surface capacitance ($\Ci$).

The bulk equations governing the electrolyte in the electrodes are
\subeq{
	\label{SM-sys:electrode_liquid}
	\begin{align}
	\pderiv{\cli}{t}&=\nabla\cdot\left(\nabla \cli +\nue^{-1}\theta\left(1+\gamma\cli\right)\nabla\Phiei\right)+\nabla\cdot\veciei, \label{SM-eqn:nd_cli}\\
	\veciei&=-\left(1-\Da\right)\nabla \cli-\nue^{-1}(1+\gamma\cli)\nabla\Phiei,\label{SM-eqn:ohme}\\
	\phiei\nabla\cdot\veciei&=\Gi\left(\gibar+\Ci\pderiv{}{t}(\Phisi-\Phiei)\right),
	\end{align}
}
and the dimensionless numbers
\begin{align}
\Da = \frac{D_A}{D_L}, \quad
\nue = \frac{i_0 L F}{R T_a \sigma_e}, \quad
\theta = \frac{\mu_{L}}{\mu_L + \mu_A}, \quad
\gamma = \frac{\Delta c}{c_{L0}},
\label{eqn:nondim_2}
\end{align}
denote the ratio of anion diffusivity to lithium diffusivity ($\Da$), the relative electical resistivity of the electrolyte ($\nue$), the relative lithium mobility ($\theta$), and the relative change in the concentration of lithium ions in the electrolyte ($\gamma$), with
\begin{align}
\sigma_e = F^2 c_{L0} (\mu_L + \mu_A)
\label{eqn:sigma_e}
\end{align}
being the (dimensional) ionic conductivity.

Finally, the bulk equations for the electrolyte in the separator are given by
\begin{subequations}\label{SM-sys:separator2}
	\begin{align}
	\pderiv{\cls}{t}&=\nabla\cdot\left(\nabla \cls+\nu_e^{-1}\theta\left(1+\gamma\cls\right)\nabla\Phies\right), \label{SM-eqn:nd_cls}\\
	\vecies&=-\left(1-\Da\right)\nabla \cls-\nue^{-1}(1+\gamma\cls)\nabla\Phies,\label{SM-eqn:ohmes}\\
	\nabla\cdot\vecies&=0.\label{SM-eqn:ohmas}
	\end{align}
\end{subequations}

The reaction-diffusion equations for the concentration of the lithium ions in the electrolyte given by \cref{SM-eqn:nd_cli} and \cref{SM-eqn:ohme} can be simplified by eliminating their dependence on the gradient in electric potential using \cref{SM-eqn:ohme,SM-eqn:ohmes}, resulting in 
\subeq{
	\begin{align}
	\pderiv{\cli}{t}&=\R \nabla^2 \cli + (1 - \theta) \nabla \cdot \veciei, \label{SM-eqn:nd_cli2}\\
	\pderiv{\cls}{t}&=\R \nabla^2\cls, \label{SM-eqn:nd_cls2}
	\end{align}
}
where $\R = 1 - \theta (1 - \Da)$ is sometimes referred to as a retarded diffusion coefficient and has applications in chemical adsorption \cite[page 422]{Fowler2011}. Equation \cref{SM-eqn:nd_cls2} confirms that diffusion is indeed the dominant mechanism of lithium transport in the separator. We will replace \cref{SM-eqn:nd_cli,SM-eqn:nd_cls} with \cref{SM-eqn:nd_cli2,SM-eqn:nd_cls2}, respectively. 

The non-dimensional overpotential becomes
\begin{align}
\etai=\Phisi-\Phiei-\mnn{U}{i}, \qquad \mnn{U}{i} = \log(\Vi). \label{SM-eqn:butleretadef}
\end{align}
The constant $\Ui$ and composition-dependent $\Vi$ contributions to the open-circuit potential are defined as
\begin{align}
\Ui=\frac{\deltai \mn{K}{L}{i}(1-\xii)}{\mn{K}{a}{i}}, \quad
\Vi=\frac{(1+\gamma\cli)[1-\deltai\xii(1-\xii)^{-1}\gamma\csi]}{1+\deltai\gamma\csi}.
\label{eqn:UVdef}
\end{align}
The dimensionless numbers
\begin{align}
\deltai = \frac{c_{L0}}{\mn{c}{a0}{i}}, \quad
\xii = \frac{\mn{c}{a0}{i}}{\csimax},
\label{eqn:nondim_3}
\end{align}
represent the different relative initial concentrations of lithium.  The ratio of solid lithium to the maximal value is defined as the state of charge and $\xii$ is the initial state of charge, a common variable for initial battery parameterisation. The non-dimensional Butler-Volmer kinetics can be written as
\begin{subequations}\label{SM-sys:nondimbutler}
	\begin{align}
	\gibar&=\ji\left(\exp\left[(1-\betai)\etai\right]-\exp\left[-\betai\etai\right]\right),\\
	\ji&=\left(1+\deltai\gamma\csi\right)^{\betai}\left(1-\deltai\xii(1-\xii)^{-1}\gamma\csi\right)^{1-\betai}(1+\gamma\cli)^{1-\betai}. \label{SM-eqn:nd_j0i}
	\end{align}
\end{subequations}

The dimensionless boundary conditions at the positive electrode-collector interface are
\subeq{
	\label{sys:BC}
	\begin{alignat}{2}
	\phisp\vecisp\cdot\vec{n}&=-\mathcal{I}, \quad &&x = 0, \label{bc:nd_current_1}\\ 
	\veciep\cdot\vec{n}&=0, \quad &&x = 0, \label{bc:nd_current_2} \\
	\nabla \clp \cdot \vec{n} &= 0, \quad &&x = 0, \label{bc:no_flux_1}\\
	\nabla \csp \cdot \vec{n} &= 0, \quad &&x = 0.
	\end{alignat}
	Similarly, the boundary conditions at the electrode-separator interfaces are
	\begin{alignat}{2}
	\cli - \cls &= 0, \quad &&x = x_p,\,x_n;\\
	(\phiei\nabla\cli-\phies\nabla\cls)\cdot\vec{n}&=0,\quad &&x = x_p,\,x_n;\\
	(\phiei\nabla\Phiei-\phies\nabla\Phies)\cdot\vec{n}&=0,\quad &&x = x_p,\,x_n;\\
	\nabla\csi\cdot\vec{n}&=0,\quad &&x = x_p,\,x_n;\label{bc:no_flux_2} \\ 
	\vecisi\cdot\vec{n}&=0,\quad &&x = x_p,\,x_n; \label{bc:nd_current_3} \\
	\Phiei-\Phies &= 0,\quad &&x = x_p,\,x_n. \label{SM-bc:phi_e_cont}
	\end{alignat}
	The conditions at the negative electrode-collector interface are
	\begin{alignat}{2}
	\Phien&=0, \quad &&x = 1,\label{SM-bc:phi_e_ground} \\ 
	\vecien\cdot\vec{n}&=0, \quad &&x = 1, \label{bc:nd_current_4} \\
	\nabla \cln \cdot \vec{n} &= 0, \quad &&x = 1, \label{bc:no_flux_3}\\
	\nabla \csn \cdot \vec{n} &= 0, \quad &&x = 1.
	\end{alignat}
}
Finally, the initial conditions are given by $\csi(\vec{x},0) = 0$, $\cli(\vec{x},0) = 0$, $\Phiei(\vec{x},0) = 0$, and $\Phisi(\vec{x},0) = 0$.

Typically, the aspect ratio satisfies $\alpha\ll1$, justifying the one-dimensional model assumption used in the main text (see \cref{sec:volumeaverage}).

\section{Implicit Solution for $\Phisi$ when $\beta=1/2$}\label{sec:etasol}
In \cref{sec:ass1}, we showed that the overpotential in the negative electrode comes from solving \cref{eqn:cap1DE} and that this could be solved analytically when $\betan=1/2$.  If we take this to be true then a first integral of \cref{eqn:cap1DE} reveals
\begin{align}
\int_0^{\Phisn}\frac{\diff u}{\gnhat-2\sinh\left(u/2\right)}=\tilde{t},\label{eqn:tauneta}
\end{align}
where $\gnhat=\tfrac{\mathcal{I}}{\Gn(1-x_n)}$.  This can be solved and simplified yielding
\begin{align}
\arctanh\left(\frac{\sqrt{\gnhat^2+4}}{\gnhat+2\exp\left(-\Phisn/2\right)}\right)-\arctanh\left(\frac{\sqrt{\gnhat^2+4}}{\gnhat+2}\right)=\frac{\sqrt{\gnhat^2+4}}{4}\tilde{t}.\label{eqn:taunexplicit}
\end{align}
The steady state for this is given by
\begin{align}
{\Phisn^*}=2\log\left(\frac{\gnhat+\sqrt{\gnhat^2+4}}{2}\right).\label{eqn:etanc}
\end{align}
Similarly if we take $\betap=1/2$ and integrate \cref{eqn:cap2gp} then
\begin{align}
\int_0^{\Phisp}\frac{\diff u}{\gnhat-2\sinh\left(u/2\right)}=\check{t},\label{eqn:taupeta}
\end{align}
where $\gphat=-\tfrac{\mathcal{I}}{\Gp x_p}$ which can also be solved to get
\begin{align}
\arctanh\left(\frac{\gphat+2\exp\left(-\Phisp/2\right)}{\sqrt{\gphat^2+4}}\right)-\arctanh\left(\frac{\gphat+2}{\sqrt{\gphat^2+4}}\right)=\frac{\sqrt{\gphat^2+4}}{4}\check{t}\label{eqn:taupexplicit}
\end{align}
with steady state
\begin{align}
{\Phisp^*}=2\log\left(\frac{\gphat+\sqrt{\gphat^2+4}}{2}\right).\label{eqn:etapc}
\end{align}
The discrepency between \cref{eqn:taunexplicit} and \cref{eqn:taupexplicit} is due to simplification of logarithms based on the sign of the argument.

\section{Parameter values}\label{sec:params}

Typical parameters for physical constants of the volume-averaged cell model \cref{sys:avgmodel} are given in \cref{tab:params1} and \cref{tab:params2}.
% Table generated by Excel2LaTeX from sheet 'Battery Params'
\begin{table}[htbp]
	\centering
	\caption{Parameters applied to the entire battery}
	\begin{tabular}{|c|c|}
		\hline
		Parameter (Units) & Value (Reference) \\
		\hline
		$H$ (m) & 65$\tten{-3}$ \cite{Li2014} \\
		\hline
		$x_p$ (m) & 70$\tten{-6}$ \cite{Li2014} \\
		\hline
		$x_n$ (m) & 95$\tten{-6}$ \cite{Li2014} \\
		\hline
		$L$ (m) & 129$\tten{-6}$ \cite{Li2014} \\
		\hline
		$A_{\rm cell}$ (m\unit{2}) & 16.94$\tten{-2}$ \cite{Li2014} \\
		\hline
		$F$ (C mol\unit{-1}) & 96487 \cite{Newman2004} \\
		\hline
		$R$ (J mol\unit{-1} K\unit{-1}) & 8.314  \cite{Newman2004} \\
		\hline
		$T_a$ (K) & 298.15 (Chosen) \\
		\hline
		$I_{\rm app}$ (A) & 2.3 \cite{nanobattery2} \\
		\hline
		$i_0$ (A m\unit{-2}) & 13.6 \cref{eqn:i0def} \\
		\hline
		$\theta$ & 0.363 (\cite{Li2014, Amiribavandpour2015}) \\
		\hline
		
	\end{tabular}%
	\label{tab:params1}%
\end{table}%

% Table generated by Excel2LaTeX from sheet 'Individual Params'
\begin{table}[htbp]
	\centering
	\begin{threeparttable}
		\caption{Physical parameters associated with the electrodes, separator, and electrolyte.}
		\begin{tabular}{|c|c|c|c|c|}
			\hline
			Parameter (Units) & \multicolumn{4}{c|}{Value (Reference)} \\
			\hline
			& Positive Electrode & Negative Electrode & Electrolyte & Separator \\
			\hline
			$\mnn{a}{i}$ (m\unit{-1}) & 3.53 $\tten{7}$ \cite{Li2014} & 4.71 $\tten{5}$ \cite{Li2014} &       &  \\
			\hline
			$\phiei$ & 0.33 \cite{Li2014} & 0.33 \cite{Li2014} &       & 0.54 \cite{Li2014} \\
			\hline
			$\phisi$ & 0.43 \cite{Li2014} & 0.55 \cite{Li2014} &       &  \\
			\hline
			$D_L$ (m\unit{2} s\unit{-1}) &       &       & 2.6 $\tten{-10}$  \cite{Amiribavandpour2015} &  \\
			\hline
			$D_A$ (m\unit{2} s\unit{-1}) &       &       & 4.56 $\tten{-10}$ \cref{eqn:mobility} &  \\
			\hline
			$\mn{D}{a}{i}$ (m\unit{2} s\unit{-1}) & 5 $\tten{-14}$ \tnote{b} & 3.9 $\tten{-14}$ \cite{Li2014} &       &  \\
			\hline
			$\mu_L$ (m\unit{2} mol J\unit{-1} s\unit{-1}) &       &       & 1.05 $\tten{-13}$  \cref{eqn:mobility} &  \\
			\hline
			$\mu_A$ (m\unit{2} mol J\unit{-1} s\unit{-1}) &       &       & 1.84 $\tten{-13}$ \tnote{a} &  \\
			\hline
			$\mn{\sigma}{a}{i}$ (S m\unit{-1}) & 2.15  \tnote{b} & 100\cite{Li2014} &       &  \\
			\hline
			$\sigma_e$ (S m\unit{-1}) &       &       & 3.23 \cref{eqn:sigma_e} &  \\
			\hline
			$\csimax$ (mol m\unit{-3}) & 22806 \cite{Li2014} & 31370 \cite{Li2014}  &       &  \\
			\hline
			$\mn{c}{a0}{i}$ (mol m\unit{-3}) & 0.022 $\cspmax$ \cite{Li2014} & 0.86 $\csnmax$ \cite{Li2014} &       &  \\
			\hline
			$c_{L0}$ (mol m\unit{-3}) &       &       & 1200 \cite{Li2014} &  \\
			\hline
			$\mnn{\beta}{i}$ & 0.5   & 0.5   &       &  \\
			\hline
			$\mnn{\hat{C}}{i}$ (m\unit{2.5} mol\unit{-0.5} s\unit{-1}) & 1.4$\tten{-12}$ \tnote{c} & 3$\tten{-11}$ \tnote{c} &       &  \\
			\hline
			$\mn{g}{0}{i}$ (A m\unit{-2}) & 1.57$\tten{-2}$ \cref{eqn:g0def} & 1.09 \cref{eqn:g0def} &       &  \\
			\hline
			$\mn{C}{\Gamma}{i}$ (F m\unit{-2}) & 0.2 \cite{Li2014} & 0.2 \cite{Li2014} &       &  \\
			\hline
		\end{tabular}%
		\begin{tablenotes}\footnotesize
			\item [a] see discussion following \cref{eqn:mobility}
			\item [b] average of values from \cite{Li2014} and \cite{Amiribavandpour2015}
			\item [c] see discussion surrounding \cref{eqn:Chats}
		\end{tablenotes}
		\label{tab:params2}%
	\end{threeparttable}
\end{table}%

Some of the parameters listed have been adapted or computed based on certain assumptions and we now outline the details of that procedure.  Firstly, we assume the Nernst-Einstein relation applies \cite{Newman2004},
\begin{align}
\mu_{i}=\frac{D_{i}}{RT_a} \label{eqn:mobility}
\end{align}
consistent with other literature \cite{Wang2012,Li2014,Amiribavandpour2015}.  The parameter $\theta$ appearing in \cref{SM-sys:electrode_liquid} is the transference number and is a measure of the efficacy of a particular ion as a carrier charge.  Using the transference number $\theta=0.363$ given by Ref.~\cite{Amiribavandpour2015} we can use the definition of $\theta$ in \cref{eqn:nondim_2} and the mobility equation \cref{eqn:mobility} to determine that the diffusivity and mobility of the anion $A$ are $D_{A}=4.56\tten{-10}$ m\unit{2} s\unit{-1} and $\mu_{A}=1.84\tten{-13}$ m\unit{2} mol J\unit{-1} s\unit{-1} respectively.  Using equation \cref{eqn:sigma_e} we get that the ionic conductivity is $\sigma_e=3.22$ S m\unit{-1} consistent with orders of magnitude in Refs.~\cite{Amiribavandpour2015, Smith2006}.

The chemical rate constants in the local current density \cref{eqn:j0} are often not provided individually but instead as a ratio or product.  For example, in Li \etal\cite{Li2014}, the product
\begin{align}
\mnn{\hat{C}}{i}=\frac{\mn{K}{a}{i}^{\betai}\mn{K}{L}{i}^{1-\betai}}{{\csimax}^{1-\betai}},\label{eqn:Chats}
\end{align}
is provided and therefore we can rearrange to determine the rate constant product required. 

For comparison to experimental data, the theoretical open-circuit potential defined by \cref{eqn:Ui} will be replaced by empirical formulae from Safari and Delacourt \cite{Safari2011b} for an ANR266450m1A battery (see Ref.~\cite{nanobattery2}). The negative electrode is graphite while the positive electrode is lithium iron phosphate. The use of empirical expressions for the open-circuit potential is often favoured in battery modelling because they capture many of the electrode phenomena such as phase change.  The empirical formula for these electrodes as a function of the state of charge, $\lambda$, is
\begin{subequations}
	\begin{align}
	&\begin{aligned}
	\mn{U}{\rm ref}{n}=~&0.6379+0.5416\exp(-305.5309\mnn{\lambda}{n})+0.044\tanh(-(\mnn{\lambda}{n}-0.1958)/0.1088)\\
	&-0.1978\tanh((\mnn{\lambda}{n}-1.0571)/0.0854)-0.6875\tanh((\mnn{\lambda}{n}+0.0117)/0.0529)\\
	&-0.0175\tanh((\mnn{\lambda}{n}-0.5692)/0.0875),
	\end{aligned}\\
	&\begin{aligned}
	\mn{U}{\rm ref}{p}=~&3.4323-0.8428\exp(-80.2493(1-\mnn{\lambda}{p})^{1.3198})\\
	&-3.2474\tten{-6}\exp(20.2645(1-\mnn{\lambda}{p})^{3.8003})\\
	&+3.2482\tten{-6}\exp(20.2646(1-\mnn{\lambda}{p})^{3.7995}),
	\end{aligned}
	\end{align}
\end{subequations}
and their plots are in \cref{fig:OCV} of the main text. We choose this OCV because it is for the same type of battery used by Li \etal in \cite{Li2014} with whom we compare our results.

\section{Numerical Details}\label{sec:numerics}
We need to simulate the model \cref{sys:electrode_solid}, \cref{sys:electrode_liquid}, and \cref{sys:separator2}.  Before discretising, we will simplify the problem by removing the explicit current dependence via \cref{eqn:ohma,eqn:ohme} for each of the electrodes and \cref{eqn:ohmas,eqn:ohmes} for the separator.  The resulting initial value problem is
\begin{subequations}\label{sys:ivp}
	\begin{align}
	\pderiv{\csi}{t}=&\Di\pderiv[2]{\csi}{x}-\frac{1}{\nusn}\pderiv[2]{\Phisi}{x}\\
	\pderiv{\psii}{t}=&\frac{1}{\Ci}\left(\frac{\phisi}{\nusn\Gi}\pderiv[2]{\Phisi}{x}-\gibar\right)\\
	\pderiv{\cli}{t}=&\mathcal{R}\pderiv[2]{\cli}{x}+\frac{\phisi(1-\theta)}{\phie\nusn}\pderiv[2]{\Phisi}{x}
	\end{align}
	for each electrode and
	\begin{align}
	\pderiv{\cls}{t}=\mathcal{R}\pderiv[2]{\cls}{x}
	\end{align}
	for the separator, where
	\begin{align}
	\psii=\Phisi-\Phiei.
	\end{align}
	There are no explicit time derivatives present for $\Phiei$ and $\Phisi$ which instead are constrained through other means.  The constraint for the solid potential is simply,
	\begin{align}
	\psii-\Phisi+\Phiei=0,
	\end{align}
	while the constraint for $\Phiei$ comes from integrating the global charge conservation \cref{eqn:phaseneutral}. Doing so and using the boundary conditions yields
	\begin{align*}
	\phisi\isi+\phie\iei=-\mathcal{I}.
	\end{align*}
	Once again eliminating currents furnishes the additional constraint on $\Phiei$,
	\begin{align}
	\mathcal{I}=&\frac{\phisi}{\nusn}\pderiv{\Phisi}{x}+\phie(1-\Da)\pderiv{\cli}{x}+\phie\nue^{-1}(1+\gamma \cli)\pderiv{\Phiei}{x}
	\end{align}
\end{subequations}
where we note that the condition for the separator excludes a solid phase potential term.  The boundary conditions for this problem are (see \cref{bc:se_fluid}-\cref{bc:dcli})
\begin{alignat}{2}
\pderiv{\csp}{x}=\pderiv{\clp}{x}=\pderiv{\Phiep}{x}=&0, &\quad x &=0, \\
\pderiv{\Phisp}{x}=&\frac{\nusp\mathcal{I}}{\phisp}, &\quad x &= 0, \\
\pderiv{\csp}{x}=\pderiv{\Phisp}{x}=[\cli]=\left[\phiei\pderiv{\cli}{x}\right]=\left[\phiei\pderiv{\Phiei}{x}\right]=&0,
&\quad x &= x_p, \\
\pderiv{\csn}{x}=\pderiv{\Phisn}{x}=[\cli]=\left[\phiei\pderiv{\cli}{x}\right]=\left[\phiei\pderiv{\Phiei}{x}\right]=&0,
&\quad  x&=x_n\\
\pderiv{\csn}{x}=\pderiv{\cln}{x}=\pderiv{\Phien}{x}=\Phien=&0, &\quad x &= 1, \\
\pderiv{\Phisn}{x}=&\frac{\nusn\mathcal{I}}{\phisn}, &\quad x &= 1,
\end{alignat}
where $[\cdot]$ is the jump across an interface.  The initial conditions are $\csi=\cli=\Phiei=\psii=\Phisi=0$.

\subsection{Domain Discretisation}
The battery problem has three domains, $\Omega_p=\{x|x\in[0,x_p]\}$, $\Omega_s=\{x|x\in[x_p,x_n]\}$, and $\Omega_n=\{x|x\in[x_n,1]\}$.  We prescribe $N$ points in each domain (we take $N=49$ in \cref{sec:comparison} of the main text) using a cell-centered grid with spacing $h_p=x_p/(N+1)$ in $\Omega_p$, $h_s=(x_n-x_s)/(N+1)$ in $\Omega_s$  and $h_n=(1-x_n)/(N+1)$ in $\Omega_n$.  We spatially discretise \cref{sys:ivp} using central differences, i.e.\ if we denote the approximation of $u(x_{k+1/2})$ by $u_{k+1/2}$ then
\begin{align}
\pderiv{u^i}{x}=&\frac{u^i_{k+3/2}-u^i_{k-1/2}}{2h_i} + \Ord{h_i^2}\\
\pderiv[2]{u^i}{x}=&\frac{u^i_{k-1/2}-2u^i_{k+1/2}+u^i_{k+3/2}}{h_i^2} + \Ord{h_i^2},
\end{align}
where ghost points are employed for values outside of the domain.  This leads to the $(N+1)\times(N+1)$ derivative, $\Dx{i}{j}{k}$, and second derivative, $\Dxx{i}{j}{k}$ matrices with subscripts $j$ and $k$ as D, N, L, or R for Dirichlet, Neumann, left-continuous, or right-continuous boundary conditions respectively.  Continuity introduces the $(N+1)\times(N+1)$ matrices $\C{i}{1}{L}$ and $\C{i}{2}{L}$ for left continuity of the first and second derivative respectively which are zero matrices except for entries in the last column of the first row.  Similarly there are right-continuity matrices $\C{i}{1}{R}$ and $\C{i}{2}{R}$ which have a non-zero entry in the first column of the last row.  Finally we define $\vec{y}=[\vec{y}_1,\vec{y}_2]^T$ with
\subeq{
	\begin{align}
	\vec{y}_i&=[\csp,\psip,\clp,\cls,\csn,\psin,\cln]^T, \\
	\vec{y}_2&=[\Phiep,\Phies,\Phien,\Phisp,\Phisn]^T
	\end{align}
}
to separate the explicit time-dependent and algebraically constrained problems.  The discrete version of \cref{sys:ivp} then becomes
\begin{align}
\begin{bmatrix}
\deriv{\vec{y}_1}{t}\\
0
\end{bmatrix}=A\vec{y}+
\begin{bmatrix}
0  &  0\\
0  &  (I+\gamma C_L)B
\end{bmatrix}
\begin{bmatrix}
\vec{y}_1\\
\vec{y}_2
\end{bmatrix}
+\vec{b}(\vec{y})
\end{align}
where $I$ is the identity matrix and $A$, $B$, and $C_L$ are defined as
\begin{align}
A = \begin{bmatrix}
\Dp\Dxxp{N}{N}	&	0	&	0	&	0	&	0	&	0	&	0	&	0	&	0	&	0	&	-\frac{1}{\nusp}\Dxxp{N}{N}	&	0\\
0	&	0	&	0	&	0	&	0	&	0	&	0	&	0	&	0	&	0	&	\frac{\phisp}{\Cp\nusp\Gp}\Dxxp{N}{N}	&	0\\
0	&	0	&	\R\Dxxp{N}{C}	&	\R\Cxxp{R}	&	0	&	0	&	0	&	0	&	0	&	0	&	\frac{(1-\theta)\phisp}{\phie\nusp}\Dxxp{N}{N}	&	0\\
0	&	0	&	\R\Cxxs{L}	&	\R\Dxxs{N}{N}	&	0	&	0	&	\R\Cxxs{R}	&	0	&	0	&	0	&	0	&	0\\
0	&	0	&	0	&	0	&	\Dn\Dxxn{N}{N}	&	0	&	0	&	0	&	0	&	0	&	0	&	-\frac{1}{\nusn}\\
0	&	0	&	0	&	0	&	0	&	0	&	0	&	0	&	0	&	0	&	0	&	\frac{\phisn}{\Cn\nusn\Gn}\\
0	&	0	&	0	&	\R\Cxxn{L}	&	0	&	0	&	\R\Dxxn{C}{N}	&	0	&	0	&	0	&	0	&	\frac{(1-\theta)\phisn}{\phie\nusn}\Dxxn{N}{N}\\
0	&	0	&	\tilde{a}\Dxp{N}{C}	&	\tilde{a}\Cxp{R}	&	0	&	0	&	0	&	0	&	0	&	0	&	\frac{\phisp}{\nusp}\Dxn{N}{N}	&	0\\
0	&	0	&	\tilde{a}\Cxs{L}	&	\tilde{a}\Dxs{C}{C}	&	0	&	0	&	\tilde{a}\Cxs{R}	&	0	&	0	&	0	&	0	&	0\\
0	&	0	&	0	&	\tilde{a}\Cxn{L}	&	0	&	0	&	\tilde{a}\Dxn{C}{N}	&	0	&	0	&	0	&	0	&	\frac{\phisn}{\nusn}\Dxn{N}{N}\\
0	&	I	&	0	&	0	&	0	&	0	&	0	&	I	&	0	&	0	&	-I	&	0\\
0	&	0	&	0	&	0	&	0	&	I	&	0	&	0	&	0	&	I	&	0	&	-I\label{eqn:Amat}
\end{bmatrix}&,\nonumber \\
B=\frac{\phie}{\nue}\begin{bmatrix}
\Dxp{N}{C}	&	\Cxp{R}	&	0	&	0	&	0\\
\Cxs{L}	&	\Dxs{C}{C}	&	\Cxs{R}	&	0	&	0\\
0	&	\Cxn{L}	&	\Dxn{C}{D}	&	0	&	0
\end{bmatrix}&, \\
C_L=\begin{bmatrix}
\textrm{diag}(\clp)	&	0	&	0\\
0	&	\textrm{diag}(\cls)	&	0\\
0	&	0	&	\textrm{diag}(\cln)
\end{bmatrix}&, \nonumber
\end{align}
respectively where $\tilde{a}=\phie(1-\Da)$.  The vector $\vec{b}$ is defined as
\begin{align}
\begin{bmatrix}
\frac{\mathcal{I}}{\nusp^2\phisp h_p^2}\vec{e}_L\\
-\frac{\mathcal{I}}{\Cp\nusp^2\Gp h_p^2}\vec{e}_L - \frac{\gpbar}{\Cp}\\
-\frac{(1-\theta)\mathcal{I}}{\phie\nusp^2 h_p^2}\vec{e}_L\\
0\\
\frac{\mathcal{I}}{\nusn^2\phisn h_n^2}\vec{e}_R\\
-\frac{\mathcal{I}}{\Cn\nusn^2\Gn h_n^2}\vec{e}_R - \frac{\gnbar}{\Cn}\\
-\frac{(1-\theta)\mathcal{I}}{\phie\nusn^2 h_n^2}\vec{e}_R\\
\frac{\mathcal{I}}{2\nusp^2}\vec{e}_L-\mathcal{I}\vec{e}\\
-\mathcal{I}\vec{e}\\
\frac{\mathcal{I}}{2\nusn^2}\vec{e}_R-\mathcal{I}\vec{e}\\
0\\
0
\end{bmatrix}
\end{align}
where $\vec{e}_L=[1,0,\dots,0]^T$, $\vec{e}_R=[0,\dots,0,1]^T$, and $\vec{e}$ is a vector of all ones.  This vector includes nonlinear terms from the the Butler-Volmer kinetics, $\gibar$. Note that in \cref{eqn:Amat} derivative and continuity matrices include the volume fraction $\phiei$ where appropriate. We note that because the source vector $\vec{b}=\vec{b}(\vec{y})$ and the matrix $C_L=C_L(\vec{y})$ that the problem is non-linear. We solve the problem using fully implicit backward Euler as a time-stepping method.
\end{appendix}

\bibliographystyle{siamplain}
\bibliography{bib}
\end{document}

%% file: battery_asymptotics_shared.tex
\usepackage{lipsum}
\usepackage{amsfonts}
\usepackage{graphicx}
\usepackage{epstopdf}
\usepackage{algorithmic}
\usepackage{mhchem}
\usepackage{threeparttable}
\usepackage{subfig}

\ifpdf
\DeclareGraphicsExtensions{.eps,.pdf,.png,.jpg}
\else
\DeclareGraphicsExtensions{.eps}
\fi

\graphicspath{{ink/},{figs/}}

\newsiamremark{remark}{Remark}
\newsiamremark{hypothesis}{Hypothesis}
\crefname{hypothesis}{Hypothesis}{Hypotheses}
\newsiamthm{claim}{Claim}

\headers{Lithium-ion batteries}{I.~R.~Moyles, M.~G.~Hennessy, T.~G.~Myers, and B.~R.~Wetton}

\title{Asymptotic reduction of a porous electrode model for lithium-ion batteries\thanks{Submitted to the editors on \today.
		\funding{This work was funded by an Irish Research Council New Foundations Grant. IRM acknowledges support from Science Foundation Ireland under grant number SFI/13/IA/1923. TGM acknowledges the support of a Ministerio de Ciencia e Innovaci{\'o}n grant MTM2017-82317-P. MGH acknowledges support from a travel grant from the Mathematics and Applications Consortium for Science and Industry from the University of Limerick. MGH and TGM have been partially funded by the CERCA Programme of the Generalitat de Catalunya and received funding from the European Union's Horizon 2020 research and innovation programme under the Marie Sk{\l}odowska-Curie grant agreement No. 707658.}}}

\author{Iain~R.~Moyles\thanks{Department of Mathematics and Statistics, York University, 4700 Keele Street, Toronto, Ontario, M3J 1P3, Canada (\email{imoyles@yorku.ca}).}
	\and Matthew~G.~Hennessy\thanks{Mathematical Institute, University of Oxford, Andrew Wiles Building, Radcliffe Observatory Quater, Woodstock Road, Oxford, OX2 6GG, United Kingdom (\email{hennessy@maths.ox.ac.uk}).}
	\and Timothy~G.~Myers\thanks{Centre de Recerca Matem\`atica, 08193 Bellaterra (Barcelona), Spain (\email{tmyers@crm.cat}).}
	\and Brian~R.~Wetton\thanks{Department of Mathematics, University of British Columbia, Vancouver, Canada (\email{wetton@math.ubc.ca}).}}

\usepackage{amsopn}

\makeatletter
\providecommand*{\diff}%
{\@ifnextchar^{\DIfF}{\DIfF^{}}}
\def\DIfF^#1{%
	\mathop{\mathrm{\mathstrut d}}%
	\nolimits^{#1}\gobblespace}
\def\gobblespace{%
	\futurelet\diffarg\opspace}
\def\opspace{%
	\let\DiffSpace\!%
	\ifx\diffarg(%
	\let\DiffSpace\relax
	\else
	\ifx\diffarg[%
	\let\DiffSpace\relax
	\else
	\ifx\diffarg\{%
	\let\DiffSpace\relax
	\fi\fi\fi\DiffSpace}

\providecommand*{\deriv}[3][]{%
	\frac{\diff^{#1}#2}{\diff #3^{#1}}}

\providecommand*{\pderiv}[3][]{%
	\frac{\partial^{#1}#2}%
	{\partial #3^{#1}}}

\renewcommand{\vec}{\boldsymbol}

\newcommand{\tten}[1]{\times 10^{#1}}
\newcommand{\Ord}[1]{\mathcal{O}\left(#1\right)}

\newcommand{\avg}[1]{\left<{#1}\right>}

\newcommand{\etal}{\emph{et al.~}}
\newcommand{\unit}[1]{$^{#1}$}

\newenvironment{subeq}[1]{
	\begin{subequations}
		#1
	\end{subequations}
}

\newcommand{\mn}[3]{#1_{#2,\text{{#3}}}}
\newcommand{\mnn}[2]{#1_{\text{{#2}}}}

\newcommand{\clp}{\mn{c}{L}{p}}
\newcommand{\cln}{\mn{c}{L}{n}}
\newcommand{\cls}{\mn{c}{L}{s}}
\newcommand{\cli}{\mn{c}{L}{i}}

\newcommand{\csp}{\mn{c}{a}{p}}
\newcommand{\csn}{\mn{c}{a}{n}}
\newcommand{\csi}{\mn{c}{a}{i}}

\newcommand{\csphat}{\mn{\hat{c}}{a}{p}}
\newcommand{\csnhat}{\mn{\hat{c}}{a}{n}}
\newcommand{\csihat}{\mn{\hat{c}}{a}{i}}

\newcommand{\csimax}{\mn{c}{a}{i}^\text{max}}
\newcommand{\csnmax}{\mn{c}{a}{n}^\text{max}}
\newcommand{\cspmax}{\mn{c}{a}{p}^\text{max}}

\newcommand{\phiep}{\mn{\phi}{e}{p}}
\newcommand{\phien}{\mn{\phi}{e}{n}}
\newcommand{\phies}{\mn{\phi}{e}{s}}
\newcommand{\phiei}{\mn{\phi}{e}{i}}

\newcommand{\phisp}{\mn{\phi}{a}{p}}
\newcommand{\phisn}{\mn{\phi}{a}{n}}
\newcommand{\phisi}{\mn{\phi}{a}{i}}

\newcommand{\phie}{\mnn{\phi}{e}}

\newcommand{\Phiep}{\mn{\Phi}{e}{p}}
\newcommand{\Phien}{\mn{\Phi}{e}{n}}
\newcommand{\Phies}{\mn{\Phi}{e}{s}}
\newcommand{\Phiei}{\mn{\Phi}{e}{i}}

\newcommand{\Phisp}{\mn{\Phi}{a}{p}}
\newcommand{\Phisn}{\mn{\Phi}{a}{n}}
\newcommand{\Phisi}{\mn{\Phi}{a}{i}}

\newcommand{\etai}{\mnn{\eta}{i}}

\newcommand{\Phisphat}{\mn{\hat{\Phi}}{a}{p}}
\newcommand{\Phisnhat}{\mn{\hat{\Phi}}{a}{n}}
\newcommand{\Phisihat}{\mn{\hat{\Phi}}{a}{i}}

\newcommand{\Up}{\mnn{\mathcal{U}}{p}}
\newcommand{\Un}{\mnn{\mathcal{U}}{n}}
\newcommand{\Ui}{\mnn{\mathcal{U}}{i}}

\newcommand{\Vi}{\mnn{\mathcal{V}}{i}}

\newcommand{\gnbar}{\mnn{g}{n}}%{\mnn{\bar{g}}{n}}
\newcommand{\gpbar}{\mnn{g}{p}}%{\mnn{\bar{g}}{p}}
\newcommand{\gibar}{\mnn{g}{i}}%{\mnn{\bar{g}}{i}}

\newcommand{\ies}{\mn{i}{e}{s}}
\newcommand{\iei}{\mn{i}{e}{i}}

\newcommand{\isp}{\mn{i}{a}{p}}
\newcommand{\isn}{\mn{i}{a}{n}}
\newcommand{\isi}{\mn{i}{a}{i}}

\newcommand{\veciep}{\mn{\vec{i}}{e}{p}}
\newcommand{\vecien}{\mn{\vec{i}}{e}{n}}
\newcommand{\vecies}{\mn{\vec{i}}{e}{s}}
\newcommand{\veciei}{\mn{\vec{i}}{e}{i}}

\newcommand{\vecisp}{\mn{\vec{i}}{a}{p}}

\newcommand{\vecisi}{\mn{\vec{i}}{a}{i}}

\newcommand{\vecNls}{\mn{\vec{N}}{L}{s}}
\newcommand{\vecNli}{\mn{\vec{N}}{L}{i}}

\newcommand{\vecNsi}{\mn{\vec{N}}{a}{i}}

\newcommand{\Cp}{\mnn{\mathcal{C}}{p}}
\newcommand{\Cn}{\mnn{\mathcal{C}}{n}}
\newcommand{\Ci}{\mnn{\mathcal{C}}{i}}

\newcommand{\Gp}{\mnn{\mathcal{G}}{p}}
\newcommand{\Gn}{\mnn{\mathcal{G}}{n}}
\newcommand{\Gi}{\mnn{\mathcal{G}}{i}}

\newcommand{\Dp}{\mnn{\mathcal{D}}{p}}
\newcommand{\Dn}{\mnn{\mathcal{D}}{n}}
\newcommand{\Da}{\mnn{\mathcal{D}}{A}}
\newcommand{\Di}{\mnn{\mathcal{D}}{i}}

\newcommand{\nue}{\mnn{\nu}{e}}
\newcommand{\nusn}{\mn{\nu}{a}{n}}
\newcommand{\nusp}{\mn{\nu}{a}{p}}
\newcommand{\nui}{\mnn{\nu}{i}}

\newcommand{\betai}{\mnn{\beta}{i}}
\newcommand{\betap}{\mnn{\beta}{p}}
\newcommand{\betan}{\mnn{\beta}{n}}

\newcommand{\ji}{\mnn{j}{i}}

\newcommand{\deltap}{\mnn{\delta}{p}}
\newcommand{\deltan}{\mnn{\delta}{n}}
\newcommand{\deltai}{\mnn{\delta}{i}}

\newcommand{\xip}{\mnn{\xi}{p}}
\newcommand{\xin}{\mnn{\xi}{n}}
\newcommand{\xii}{\mnn{\xi}{i}}

\newcommand{\gnhat}{\hat{g}_n}%{\frac{\mathcal{I}}{\Gn(1-x_n)}}
\newcommand{\gphat}{\hat{g}_p}%{\frac{\mathcal{I}}{\Gp x_p}}

\newcommand{\psii}{\mnn{\psi}{i}}
\newcommand{\psip}{\mnn{\psi}{p}}
\newcommand{\psin}{\mnn{\psi}{n}}

\newcommand{\Dx}[3]{D^{\text{#1}}_{1,\text{{#2}{#3}}}}
\newcommand{\Dxp}[2]{\Dx{p}{#1}{#2}}
\newcommand{\Dxs}[2]{\Dx{s}{#1}{#2}}
\newcommand{\Dxn}[2]{\Dx{n}{#1}{#2}}
\newcommand{\Dxx}[3]{D^{\text{#1}}_{2,\text{{#2}{#3}}}}
\newcommand{\Dxxp}[2]{\Dxx{p}{#1}{#2}}
\newcommand{\Dxxs}[2]{\Dxx{s}{#1}{#2}}
\newcommand{\Dxxn}[2]{\Dxx{n}{#1}{#2}}

\newcommand{\C}[3]{C^{\text{#1}}_{\text{{#2},{#3}}}}
\newcommand{\Cxp}[1]{\C{p}{1}{#1}}
\newcommand{\Cxs}[1]{\C{s}{1}{#1}}
\newcommand{\Cxn}[1]{\C{n}{1}{#1}}
\newcommand{\Cxxp}[1]{\C{p}{2}{#1}}
\newcommand{\Cxxs}[1]{\C{s}{2}{#1}}
\newcommand{\Cxxn}[1]{\C{n}{2}{#1}}

\newcommand{\R}{\mathcal{R}}

\newcommand{\cnc}{-\frac{\mathcal{I}}{\phisn(1-x_n)}}
\newcommand{\cpc}{\frac{\mathcal{I}}{\phisp x_p}}

\newcommand{\fshat}[1]{\frac{(1-\theta)\mathcal{I}}{#1 \R\phie}}

\DeclareMathOperator{\arctanh}{arctanh}